\documentclass[conference]{IEEEtran}
\IEEEoverridecommandlockouts

\usepackage{graphicx}
\usepackage{booktabs} 

\graphicspath{{figure/}{figures/}}

\usepackage{color}
\usepackage{multirow}
\usepackage{subcaption}
\usepackage{tabularx}
\usepackage{graphicx}

\usepackage{eso-pic,xcolor}
\usepackage{caption}
\usepackage{graphicx}
\usepackage{verbatim}
\usepackage{subcaption}
\usepackage{multirow}
\usepackage{array}

\usepackage{enumitem}
\usepackage{siunitx}
\usepackage{amsmath}
\usepackage{amssymb}
\usepackage{mathtools}
\usepackage{tabulary}
\usepackage[bookmarks=false]{hyperref}
\usepackage{tikz}
\usepackage{lipsum}

\newcommand{\TODO}[1]{}
\renewcommand{\TODO}[1]{{\color{red} TODO: {#1}}}

\usepackage{etoolbox}
\usepackage{xstring}
\DeclareListParser{\doslashlist}{/}
\newcounter{ndnNameComponentCounter}%
\newcommand{\name}[1]{{%
  \setcounter{ndnNameComponentCounter}{0}%
  \renewcommand{\do}[1]{{%
    \ifnumgreater{\value{ndnNameComponentCounter}}{0}{\allowbreak/}{}%
    \ifnumodd{\value{ndnNameComponentCounter}}{}{}%
    \detokenize{##1}}%
    \stepcounter{ndnNameComponentCounter}}%
``{\fontfamily{cmtt}\small\selectfont\IfBeginWith{#1}{/}{/}{}\doslashlist{#1}}''%
}}

\newcommand\floor[1]{\lfloor#1\rfloor}

\newcommand\copyrighttext{%
  \footnotesize \center {\color{red} This paper was accepted for publication at the 40th IEEE International Conference on Distributed Computing Systems (ICDCS) (\url{https://icdcs2020.sg}). The copyright is with the IEEE}.}
\newcommand\copyrightnotice{%
\begin{tikzpicture}[remember picture,overlay]
\node[anchor=south,yshift=10pt] at (current page.south) {\fbox{\parbox{\dimexpr\textwidth-\fboxsep-\fboxrule\relax}{\copyrighttext}}};
\end{tikzpicture}%
}

\begin{document}
\title{DAPES: Named Data for Off-the-Grid File Sharing with Peer-to-Peer Interactions}

\author{
\IEEEauthorblockN{Spyridon Mastorakis}
\IEEEauthorblockA{University of Nebraska Omaha \\
smastorakis@unomaha.edu}
\and
\IEEEauthorblockN{Tianxiang Li}
\IEEEauthorblockA{UCLA\\
tianxiang@cs.ucla.edu}
\and
\IEEEauthorblockN{Lixia Zhang}
\IEEEauthorblockA{UCLA \\
lixia@cs.ucla.edu}}





\maketitle

\begin{abstract}

This paper introduces DAta-centric Peer-to-peer filE Sharing (DAPES), a data sharing protocol for scenarios with intermittent connectivity and user mobility. DAPES provides a set of semantically meaningful hierarchical naming abstractions that facilitate the exchange of file collections via local connectivity. This enables peers to ``make the most'' out of the limited connection time with other peers by maximizing the utility of individual transmissions to provide data missing by most connected peers. DAPES runs on top of Named-Data Networking (NDN) and extends NDN's data-centric network layer abstractions to achieve communication over multiple wireless hops through an adaptive hop-by-hop forwarding/suppression mechanism. We have evaluated DAPES through real-world experiments in an outdoor campus setting and extensive simulations. Our results demonstrate that DAPES achieves 50-71\% lower overheads and 15-33\% lower file sharing delays compared to file sharing solutions that rely on IP-based mobile ad-hoc routing.

\end{abstract}



\copyrightnotice

\section{Introduction}
\label{intro}

``Off-the-grid'' communication includes scenarios, where Internet connectivity may not be available, since the backbone infrastructure may be damaged (e.g., disaster recovery) or absent (e.g., battlefield, rural areas). Data sharing in such scenarios is vital for the dissemination of critical information (e.g., disaster status) and needs to be done through local network connectivity among the communicating entities. The communicating entities may also be mobile with intermittent connectivity to each other and the network topology may be dynamic, introducing new challenges to data sharing.

Although the communicating entities are inherently interested in the data to share, existing solutions that run on top of the IP-based network architecture~\cite{krifa2009bithoc, pucha2004ekta} typically rely on Mobile Ad-hoc Networking (MANET) routing protocols, such as DSDV~\cite{perkins1994highly} and AODV~\cite{perkins2003ad}, to establish reachability to the IP address of each entity. After that, the actual data delivery can begin. 
Moreover, in off-the-grid scenarios, IP address configuration becomes a challenge; a number of existing solutions have been proposed~\cite{nesargi2002manetconf, mcauley2000self, misra2001autoconfiguration}, which share the goal of assigning an IP address to each entity that does not collide with others. That is, in the context of off-the-grid communication, IP addresses are merely unique node identifiers, since the node location may constantly change.

In this paper, we argue that a data-centric approach to off-the-grid file sharing aligns with the objective of the communicating entities, namely the inherent interest in the data they would like to share. In line with this assertion, we propose DAta-centric Peer-to-peer filE Sharing (DAPES), which defines semantically meaningful hierarchical naming abstractions that identify the shared data directly. These names are independent of the location of the entity that produced the data or the underlying connectivity. Through these semantically meaningful names, DAPES also conveys compactly encoded information about the data that the participants of the file sharing process, called \emph{peers}, have and facilitates transmission prioritization among peers for efficient data sharing.

DAPES runs on top of Named Data Networking (NDN)~\cite{zhang2014named}, which provides a request/response communication model, directly utilizing the names defined by DAPES. DAPES leverages NDN's cryptographic primitives that bind the content of each network layer packet to its name, enabling peers to reason about data provenance and integrity. DAPES extends the NDN data-centric network layer abstractions to make use of any and all the means of connectivity, being able to fetch data from any peer that can provide it in the network. As a result, a ``traditional'' MANET routing protocol for communication across multiple wireless hops is no longer needed. Expressing the DAPES operations through semantically meaningful names, used directly by the underlying network, enables peers to make forwarding decisions based on what data is available through multiple hops over time.



The contributions of our work are the following: 

\begin{itemize}[leftmargin=0cm,itemindent=.3cm,labelwidth=\itemindent,labelsep=0cm,align=left]


\item We propose and design DAPES, a data-centric protocol for peer-to-peer file sharing in off-the-grid communication scenarios. DAPES offers unified mechanisms to maximize the utility of transmissions and mitigate collisions due to simultaneous transmissions (Section~\ref{sec:designcomponents}). As a result, peers ``make the most'' out of each (short-lived) encounter with others, minimizing the number of required transmissions. DAPES also extends NDN's data-centric forwarding plane to build short-lived knowledge about the data available around peers. In this way, multi-hop communication is achieved through an adaptive hop-by-hop forwarding/suppression mechanism (Section~\ref{sec:extensions}).



\item We implement a DAPES prototype, which we evaluate through real-world experiments in an outdoor campus setting and extensive simulations (Section~\ref{sec:evaluation}). Our results demonstrate that DAPES achieves 50-71\% lower overheads and 15-33\% lower file sharing delays than IP-based solutions that rely on MANET routing.

\end {itemize}


To the best of our knowledge, DAPES is one of the very first efforts to offer a concrete design and implementation of a data sharing solution in dynamic off-the-grid setups on top of a data-centric network substrate.

\section{Background \& Prior Work}
\label{sec:background}

In this section, we present an overview of the NDN architecture, prior related work to discuss how DAPES is inspired, but also differs from prior efforts, and a sample use-case that we use to elaborate on the DAPES design throughout the paper.

\subsection{NDN Overview}
\label{ndn-overview}

In NDN, each data packet is assigned a unique name at the time of its production. This name is used as the data identifier by the network layer and is independent of the underlying network connectivity. The NDN communication paradigm is receiver-driven; \emph{data consumers} send requests, called Interest packets, for named data packets generated by \emph{data producers}. Data names are semantically meaningful, hierarchically structured and can contain a variable number of components. For example, a consumer sends an Interest with a name \name{/cnn/daily-news/headlines} to fetch the headlines of the daily news from CNN. NDN builds communication security directly into the network architecture, since data producers cryptographically sign each data packet at the time of generation. The signature binds the content of a data packet to its name, so that a consumer can authenticate the data directly using the producer's public key~\cite{ndn-packet-format}. 

\begin{figure}[h]
	\centering
	\includegraphics[width=8.5cm]{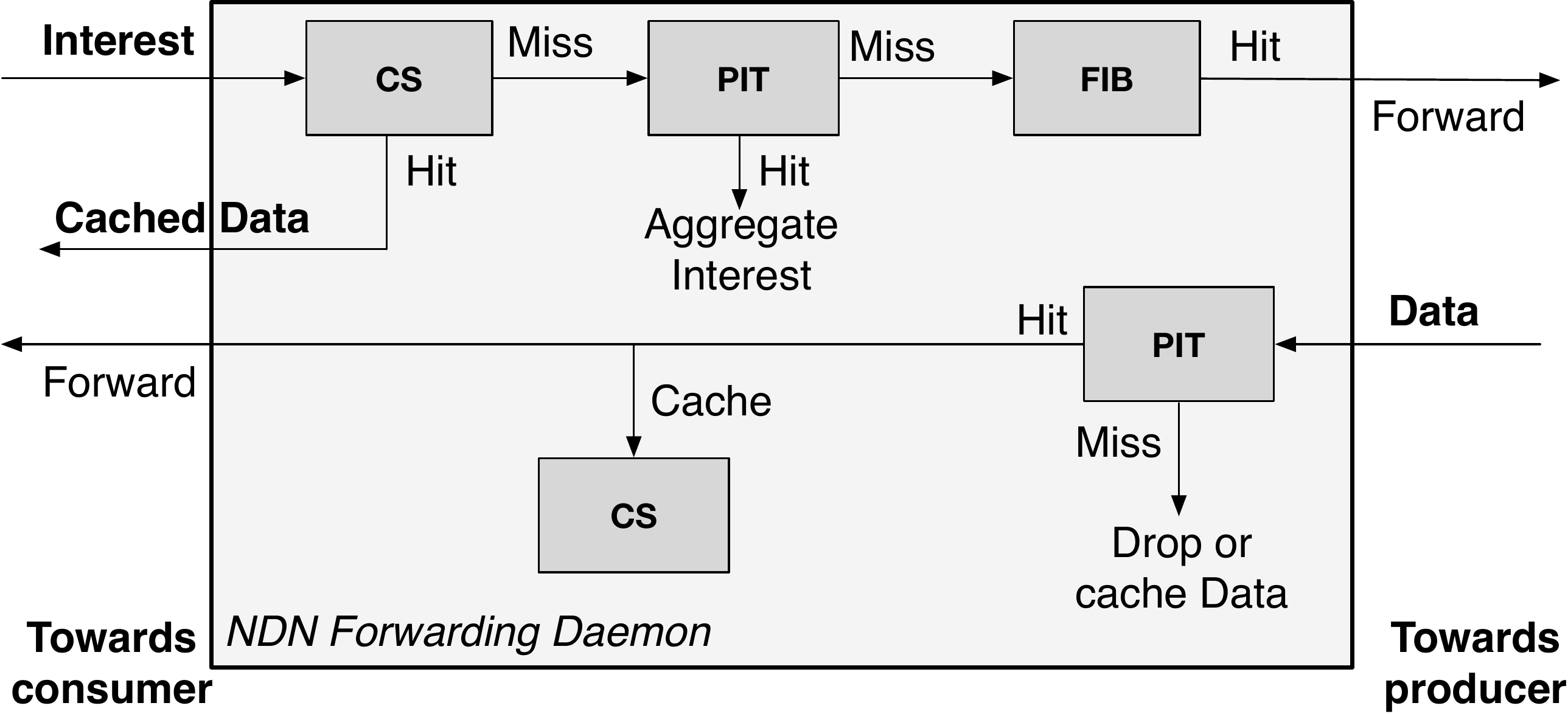}
	\caption{Packet processing by NFD}
	\label{ndn-node}
	\vspace{-1cm}
\end{figure}

NDN Forwarding Daemons (NFDs)~\cite{nfd-dev} can cache received data packets to satisfy future requests for the same data, given that each data packet is named and secured directly at the network layer. When an NFD receives an Interest, it first checks whether the requested data exists in its local Content Store (CS), as illustrated in Figure~\ref{ndn-node}. If no cached data is found, the Interest is checked against the entries of the Pending Interest Table (PIT), where state is maintained about the Interests that have been forwarded, but the corresponding data has not been received yet. If a pending Interest with the same name exists in PIT, no further forwarding is performed, since data is expected to be received. If no matching Interest is found, NFD determines how to forward the Interest based on a Longest Prefix Match (LPM) between the Interest name and the entries in its Forwarding Information Base (FIB).  A data packet uses the state in PIT, created by the corresponding Interest at each-hop NFD, to follow the reverse path back to the requesting consumer(s). 

\subsection {Prior Related Work}

\noindent \textbf{Prior work in IP:} BitTorrent~\cite{cohen2003incentives} is the most popular peer-to-peer file sharing application, focusing on infrastructure IP-based networks. Efforts to adapt BitTorrent to work in MANET~\cite{rajagopalan2006cross, gaddam2009study, sbai2008bithoc, krifa2009bithoc} largely rely on MANET routing~\cite{he2002destination, johnson2001dsr, perkins2003ad} for path discovery and maintenance between peers.  Mobility introduces additional challenges, since established paths break and new ones have to be established. Previous work~\cite{nandan2005co, lee2006code, klemm2003special, pucha2004ekta, grossglauser2006locating} has advocated that alternative solutions for multi-hop communication need to be explored (e.g., application-layer gossiping, network coding, link-layer flooding). Security is not considered in the routing and the data sharing process~\cite{yang2004security}, while IP address configuration in such infrastructure-free environments is a challenge on its own~\cite{nesargi2002manetconf, mcauley2000self, misra2001autoconfiguration}.




\noindent \textbf {Prior work in NDN:} nTorrent~\cite{mastorakis2017ntorrent}, inspired by BitTorrent, is an NDN-native application for peer-to-peer file sharing in infrastructure networks. 
Detti et al.~\cite{detti2013peer, detti2015mobile} and Malabocchia et al.~\cite{malabocchia2015using} studied the design of peer-to-peer video streaming. Prior work has also studied the design of general purpose architectures and forwarding mechanisms~\cite{amadeo2015forwarding} for MANET in NDN. E-CHANET~\cite{amadeo2013chanet} is such an architecture that runs on top of IEEE 802.11 and provides stable paths and reliable transport functions towards named data. Varvello et al.~\cite{varvello2011design} and Meisel et al.~\cite{meisel2010ad} performed an initial exploration of the design space for MANET challenges in NDN, such as resource discovery and multi-hop forwarding. Finally, Li et al.~\cite{li2019distributed} designed and implemented DDSN, a protocol for distributed dataset synchronization under disruptive network conditions.



\noindent \textbf {How DAPES is inspired and how it differs from prior work:} DAPES builds on prior work on peer-to-peer file sharing. BitTorrent uses a torrent-file that contains metadata about the shared file collection (e.g., tracker IP address, cryptographic hash of data for integrity verification), helping peers initialize their file downloading process. In a similar manner, nTorrent uses a metadata file that contains the names and the hashes of the data to request. DAPES, inspired by BitTorrent and nTorrent, uses cryptographically signed metadata (Section~\ref{subsec:torrent-file}) to help peers learn the names of the data to request and verify its integrity. BitTorrent peers use a bitmap to keep track of the data they have and leverage the ``Rarest Piece First'' (RPF) strategy to replicate data. DAPES peers also use a bitmap to encode the data peers have in a compressed manner. We explore different ways for DAPES peers to advertise this information in order to increase the efficiency of the data sharing process under intermittent connectivity (Section~\ref{subsec:bitmap}). We also propose variations of the RPF strategy, which are specifically designed to maximize the replication of rare data in dynamic communication scenarios (Section~\ref{subsec:strategy}).

Solutions for distributed dataset synchronization, such as DDSN~\cite{li2019distributed}, focus on the exchange of dynamic content, contrary to DAPES that focuses on the exchange of static content among peers. Preliminary design space explorations~\cite{meisel2010ad, varvello2011design} did not result in concrete protocol designs and solutions, while frameworks such as E-CHANET~\cite{amadeo2013chanet} did not fully exploit the data-centricity of the underlying NDN architecture to achieve their goals.


DAPES is a concrete data-centric framework for peer-to-peer file sharing in off-the-grid scenarios. DAPES functions are achieved 
through a set of mechanisms that maximize the utility of each single transmission (Sections~\ref{subsec:bitmap} and~\ref{subsec:prioritization}). At the same time, DAPES mitigates collisions due to simultaneous peer transmissions, facilitating file sharing in cases of encounters among multiple peers (Section~\ref{subsec:prioritization}). Thanks to all these mechanisms, peers can ``make the most'' out of each (potentially short-lived) encounter with others, maximizing the efficiency of the file sharing process. Furthermore, DAPES extends the underlying data-centric forwarding plane to identify what data is available over multiple wireless hops, thus making accurate forwarding/suppression decisions (Section~\ref{sec:extensions}). 


\subsection {Example Use-Case}
\label{subsec:scenario}

\begin{figure}[h]
	\centering
	\includegraphics[width=\columnwidth]{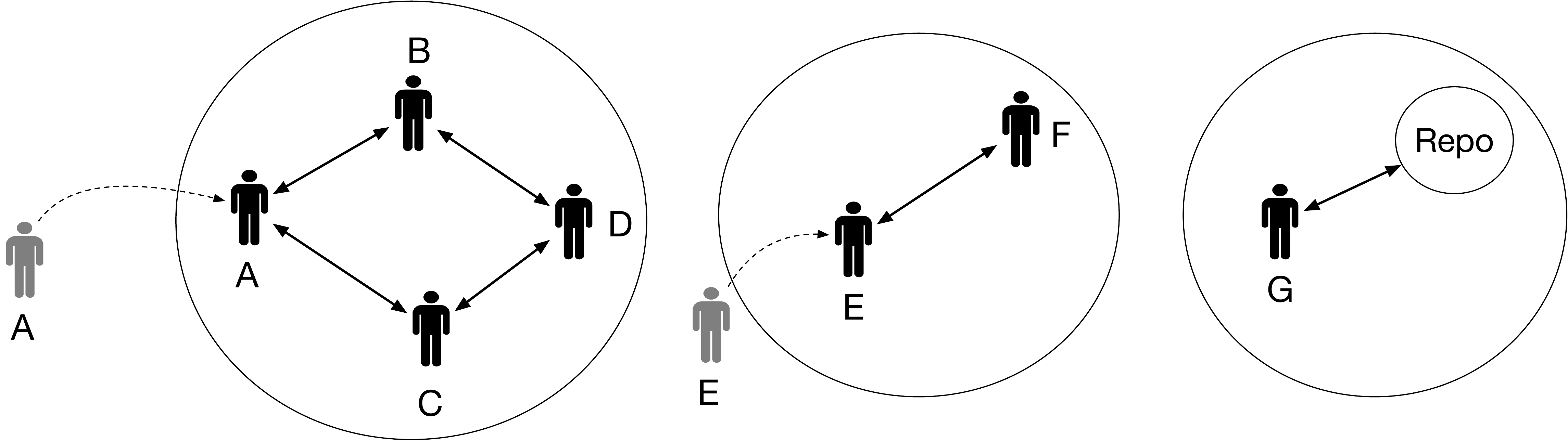}
	\caption{Example of an off-the-grid communication scenario}
	\label{off-the-grid-scenario}
	\vspace{-0.5cm}
\end{figure}

Our example use-case (Figure~\ref{off-the-grid-scenario}) assumes a rural area, where residents would like to share with other residents information about damaged parts of the infrastructure (e.g., a damaged bridge). This information needs to be resiliently, securely, and efficiently shared with as few transmissions as possible (i.e., minimal overhead and energy consumption), given that the resident devices may have limited battery power (e.g., mobile phones, tablets). We assume that residents have an instance of the DAPES application running on their device, which names individual files, segments a file into network-layer data packets and signs these packets\footnote{We assume that each resident has a pair of public and private keys to sign the packets that it generates.}, groups individual files together to create \emph{a collection of files}, and shares files with others.


Let us assume that a resident takes a picture of a damaged bridge and, through the DAPES application, names this picture (file) as \name{bridge-picture}. The resident also creates another file that contains information about the bridge location (e.g., longitude, latitude, description of surroundings) and names this file as \name{bridge-location}. Finally, the resident, through the DAPES application, creates a file collection consisting of these two files with a name \name{/damaged-bridge-1533783192} that includes the unix-timestamp of when the collection was generated. Each file in the collection consists of a number of individual data packets signed by the private key of the resident, who acts as the collection producer and starts disseminating the file collection data, aiming to notify other residents about the damage.

The data is disseminated through: (i) peer-to-peer interactions among residents as they move around in the rural area (e.g., in Figure~\ref{off-the-grid-scenario}, resident A encounters residents B, C, and D as A moves around in the rural area, while resident E encounters F), and (ii) (stationary) data repositories (``repos'' for short) locally deployed (e.g., in a rest area) to enhance data availability through collecting and serving data from/to residents~\cite{muscariello2011bandwidth} (e.g., in Figure~\ref{off-the-grid-scenario}, resident G is at a rest area, where a repo has been deployed).

\section{DAPES Design Overview}
\label{sec:design}




Our design aims to achieve efficient data sharing with low overhead among multiple peers under dynamic and adverse network conditions through: (i) a semantically meaningful and hierarchical namespace (Section~\ref{namespace}) directly utilized by an underlying data-centric network substrate, (ii) secure initialization of the data sharing process through cryptographically signed metadata (Section~\ref{subsec:torrent-file}), (iii) compact encoding of what collection data peers have (Section~\ref{subsec:bitmap}), and (iv) mechanisms for efficient data discovery, sharing, and collision mitigation (Sections~\ref{subsec:peerdisc},~\ref{subsec:strategy}, and~\ref{subsec:prioritization} respectively).

\begin{figure}[h]
  \centering
  \includegraphics[scale=0.5]{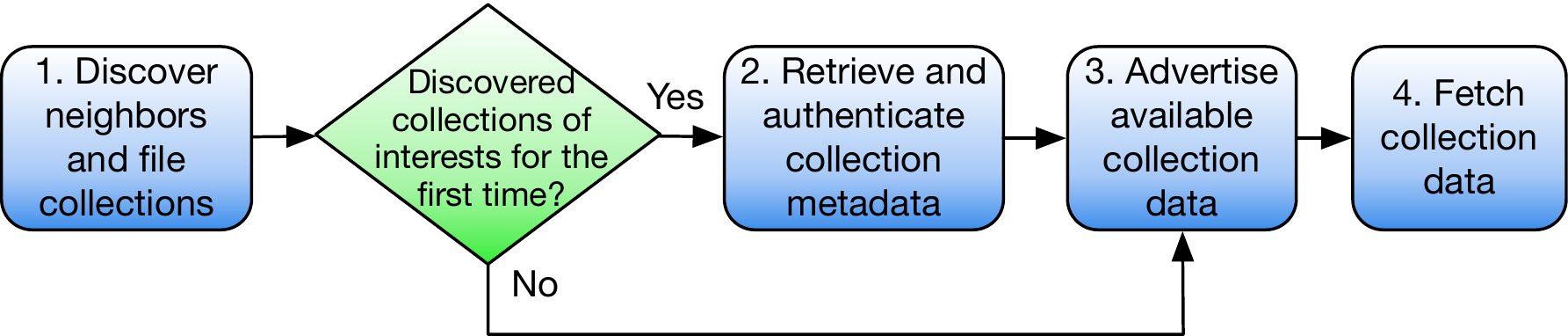}
  \caption{DAPES design overview}
  \label{Figure:protocol-overview}
  \vspace{-0.5cm}
\end{figure}

Given that peers may constantly move, a mechanism is needed to discover when peers are within the communication range of each other and what file collections they have (step 1 in Figure~\ref{Figure:protocol-overview}). Peers need to securely initialize their data sharing process by: (i) authenticating that the file collection producer can be trusted, and (ii) learning the names of the data to request in order to retrieve a file collection and verifying the integrity of the retrieved data packets. To achieve that, the file collection producer generates and signs a metadata file for the collection. When peers discover for the first time a file collection of interest through an encountered peer, they retrieve and authenticate the collection metadata (step 2 in Figure~\ref{Figure:protocol-overview}). To verify the authenticity of others, including the producer of the file collection, we assume that peers have common ``local'' trust anchors (e.g., among the residents of the rural area) established~\cite{zhang2018overview}. Based on these common trust anchors, peers verify the metadata signature and decide whether they trust the collection producer.

To reduce bandwidth consumption and communication delay, peers exchange compactly encoded information about the data they have, called ``data advertisements'' (step 3 in Figure~\ref{Figure:protocol-overview}). They prioritize the retrieval of rare data in the context of off-the-grid communication (i.e., data missing by most peers around them) through variations of the basic RPF strategy (step 4 in Figure~\ref{Figure:protocol-overview}), and use a random timer for collection data transmissions to avoid collisions. To ensure that peers are aware of as many of the available packets as possible within their communication range, they make use of a prioritization scheme for data advertisement transmissions. For MAC layer communication, peers use IEEE 802.11 in ad-hoc mode under the same SSID and channel number~\cite{crow1997ieee}.

\noindent \textbf{Communication over multiple wireless hops:} In addition to maximizing the data sharing benefits across a single wireless hop, DAPES achieves low overhead communication over multiple wireless hops. DAPES is able to make dynamic Interest forwarding/suppression decisions by assessing whether a forwarded Interest is likely to bring data back. To achieve that, peers keep track of the available data around them. When peers speculate that forwarding a received Interest will not retrieve the requested data, they suppress the Interest, while they forward received Interests when they deem that the requested data may be available around them (Section~\ref{sec:extensions}).




\section {DAPES Design Components}
\label{sec:designcomponents}

In this section, we present the components of the DAPES design in detail.

\subsection{Namespace Design}
\label{namespace}

Our goal is to enable peers to easily identify specific data of interest in a data collection (e.g., peers might not be interested in all, but only specific files in a collection). Therefore, we design a semantically meaningful and hierarchical namespace that enables peers to identify the name of a file collection, an individual file in a collection, and a data packet in a file. We identify individual packets through a sequence number, which allows us to compactly encode information about which packets in each file peers have, as we explain in Section~\ref{subsec:bitmap}. In our sample use-case (Section~\ref{subsec:scenario}), the collection of files has a name \name{/damaged-bridge-1533783192}, which includes a unix-timestamp that specifies when the collection was generated. The first packet in the first file (picture with name \name{bridge-picture}) has a name \name{/damaged-bridge-1533783192/bridge-picture/0}.



\subsection{Peer \& File Collection Discovery}
\label{subsec:peerdisc}

Given that peers are mobile and their position may constantly change, we need a mechanism to make them aware of when others are within their communication range and what collections they have. To achieve that, each peer periodically broadcasts signaling Interests, called \emph{discovery Interests}. To mitigate the overhead caused by short periods of signaling Interests, peers dynamically adjust their transmission period. Peers broadcast signaling Interests more frequently when they have recently encountered others, while their transmissions become less frequent when they are in isolation from others. Peers, receiving a discovery Interest, send a \emph{discovery data packet} back that contains the name of the metadata files for the file collections they have\footnote{The sender of a discovery data packet can learn the name of the metadata files of peers in its neighborhood by sending its own discovery Interest.}. In this way, peers can discover the file collections that others can offer. The discovery namespace consists of the application name and the name component ``discovery'' (\name{/dapes/discovery}).

This mechanism is implemented at the application layer. We aim to run on top of the existing IEEE 802.11 in ad-hoc mode, which incorporates a beaconing mechanism for clock synchronization and the discovery of neighbors, however, it does not provide any feedback to the upper layers of the network architecture~\cite{crow1997ieee}. In the example of Figure~\ref{off-the-grid-scenario}, each peer periodically broadcasts discovery Interests, which helps F and E discover each other and the collections they have after E moves within the communication range of F.

\subsection{Metadata For Secure Initialization}
\label{subsec:torrent-file}

\begin{figure}[h]
  \centering
  \includegraphics[scale=0.4]{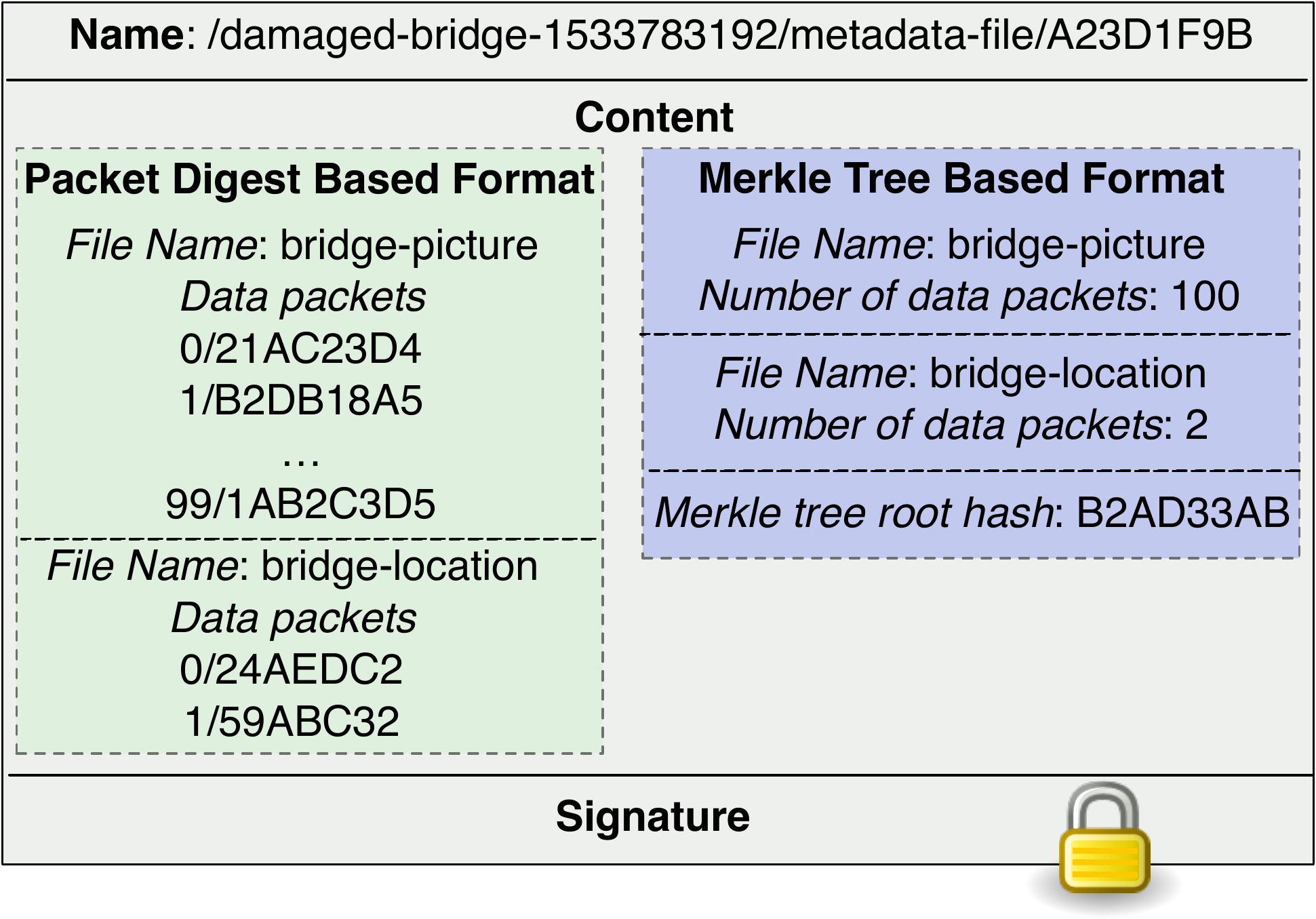}
   \vspace{-0.2cm}
  \caption{Metadata file format example}
  \label{Figure:torrent-file}
  \vspace{-0.6cm}
\end{figure}

When peers discover for the first time a file collection of interest through an encountered peer, they need to securely initialize their data sharing process. To achieve that, peers retrieve and authenticate the metadata file, which consists of one or more data packets generated and signed by the collection producer. The metadata also helps peers to discover the data namespace (name of files and individual packets) and verify the integrity of each data packet in the file collection, without having to verify its signature, which would be computationally more expensive. In the rest of this section, we present alternative metadata encoding formats (Figure~\ref{Figure:torrent-file}) and how each one achieves the above goals. These encodings involve a trade-off between the size of the metadata and how soon the integrity of each received packet can be verified.

\noindent \textbf{Packet digest based format:} Based on our hierarchical namespace, all the files in a collection and the packets in an individual file share a common name prefix. As a result, the metadata file can contain a list of ``subnames'' in the form of \name{[packet-index]/[packet-digest]} for each individual file. The subname refers to the index of a specific packet in the file followed by the packet's digest. To construct each packet's name, a peer first appends each subname to the name of the corresponding file, and then appends the resulting name to the collection name. Each subname's digest enables peers to verify the integrity of a packet as soon as it is received. This approach can result in large metadata files that need to be segmented into multiple network layer packets. Given that peers might have limited connection time during an encounter, they might need multiple encounters to fetch the entire metadata file.

\noindent \textbf{Merkle tree based format:} Peers verify the integrity of the packets through a Merkle tree~\cite{merkle1987digital}, whose root hash is generated by the collection producer and is included in the metadata content. There can also be multiple Merkle trees (e.g., one per individual file in the collection), thus, multiple root hashes can be included in the metadata. The metadata can typically fit into a single network layer packet, but all the packets in a tree need to be retrieved before their integrity can be verified. To construct each packet's name, peers first append the packet index (e.g., for a file with 100 packets, the index of the first packet is 0 and of the last packet is 99) to the name of the corresponding file, and then append the resulting name to the collection name.

\subsection{Data Advertisements}
\label{subsec:bitmap}

Given that encounters among peers might be short-lived, peers need to advertise what data they have for a collection in a compact way. Since data is hierarchically and sequentially named, we take advantage of a \emph{bitmap} data structure. Each bit refers to an individual packet, having a value of 1 if the peer has this packet and 0 if this packet is missing. After a peer downloads the collection metadata, it creates a bitmap of 0s for this collection by ordering the data packets based on the relative position of the files in the metadata and the position of the packets in each file. For the metadata of Figure~\ref{Figure:torrent-file}, the first bit of the bitmap refers to the first packet of the first file (\name{bridge-picture}), the second bit to the second packet of this file, etc. The first packet of the second file (\name{bridge-location}) corresponds to the 101st bit of the bitmap, and the second to its 102nd bit. 

After discovering the file collections an encountered peer has, peers send an Interest, called a \emph{bitmap Interest}, for each collection they are interested in. Each such Interest carries the sender's bitmap for the corresponding collection. A peer receiving such an Interest sends its own bitmap back in the content of a \emph{bitmap data packet}. 
In the example of Figure~\ref{off-the-grid-scenario}, peer E receives F's discovery data and sends a bitmap Interest carrying its bitmap. F receives E's bitmap Interest and sends back a bitmap data packet containing its own bitmap. 

\noindent \textbf{Encounters among multiple peers:} When multiple peers meet each other (e.g., A moves into the communication range of B, C and D in Figure~\ref{off-the-grid-scenario}), peers can fetch the bitmap of only some or all the other peers within their communication range. They may also select to first exchange bitmaps and then start exchanging data or interleave bitmap and data exchanges. These decisions involve a trade-off between: (i) the number of bitmaps peers retrieve, which indicates the knowledge they have about the available data (the more knowledge they have, the more efficient the downloading process will be), and (ii) how much time they have to download data. 

\noindent \textbf{Analysis:} Let us assume that peers are connected for a time interval $\Delta t$ and the transmission delay is $d$. Let T\textsubscript{delay} be the average delay for a peer to successfully transmit a bitmap (we further elaborate on T\textsubscript{delay} in Section~\ref{subsec:prioritization}). If peers exchange $b$ bitmaps before they start fetching data, the average time interval T\textsubscript{data} they will have for data fetching is:
\vspace{-0.1cm}
\begin {equation*}
     T\textsubscript{data}  =
\begin{cases}
    \Delta t - (T\textsubscript{delay} + d) * b,& \text{if }(T\textsubscript{delay} + d) * b < \Delta t \\
    0,              & \text{if } (T\textsubscript{delay} + d) * b\geq \Delta t
\end{cases}
\vspace{-0.1cm}
\end {equation*}

This equation shows that peers will have time for data fetching only if their encounter lasts more than the time they need for bitmap exchanges. When peers interleave their bitmap and data exchanges, after they fetch the first bitmap, they have an equal chance of sending a bitmap Interest or an Interest for data until $b$ bitmaps are exchanged. Assuming that $0 \leq b \leq \floor{\frac{\Delta t}{T\textsubscript{delay} + d}}$, the average time interval T\textsubscript{data} that peers have for data fetching is: 

\vspace{-0.1cm}
\begin {equation*}
     T\textsubscript{data}  =
\begin{cases}
    \Delta t - (T\textsubscript{delay} + d) * b,& \text{if }T\textsubscript{delay} + d < \Delta t \\
    0,              & \text{if } T\textsubscript{delay} + d \geq \Delta t
\end{cases}
\vspace{-0.1cm}
\end {equation*}

This equation indicates that peers will not have time for data fetching only for very small connections (i.e., they do not have enough time for a single bitmap exchange).


\subsection{Data Fetching Strategy}
\label{subsec:strategy}

After exchanging advertisements, peers know what collection data is available around them and proceed to downloading. Given that peer encounters might be short-lived and the connectivity intermittent, we need to ensure that the utility of each single data transmission is maximized by prioritizing the exchange of data missing by most peers. To achieve that, we propose two variants of the RPF strategy: (i) RPF across a peer's communication range (local neighborhood), and (ii) RPF based on the history of peer encounters. Note that this DAPES component is generic and supports the deployment of \emph{any} data fetching strategy. 

\noindent \textbf{Local neighborhood RPF:} It estimates the rarity of each packet based on how many peers within the local neighborhood do not have this data. When two or more peers exchange their bitmaps, each of them computes the rarity of each packet based on how many of the received bitmaps show a packet as missing. Each peer then creates a list of missing packets; on the top of the list are packets with higher rarity value, which will be requested first. This list is specific to each set of connected peers, and expires after the peers get disconnected, thus no long term state is maintained.

\noindent \textbf{Encounter-based RPF:} It estimates the rarity of each packet based on the history of encountered peers in the swarm, providing an estimation of how many peers in the swarm do not have each packet. Peers maintain a list of the bitmap that each encountered peer has for a certain number of encounters. Whenever a peer encounters others, it updates the list to reflect the received bitmaps. The rarity of each packet is estimated based on all the bitmaps in the list. Peers request the packets with the highest rarity values first (i.e., packets that most of the bitmaps in the list show as missing).

\noindent \textbf{Trade-offs:} The two approaches offer different ways to estimate how rare a packet is. The first one allows peers to fetch data needed by as many neighbors as possible, reducing the overall number of transmissions, without requiring peers to store long term state. The second approach prioritizes data based on peers in the swarm as a whole and requires peers to store and manage local state across multiple encounters. 

\subsection{Data Advertisement Prioritization \& Collision Mitigation}
\label{subsec:prioritization}

The efficiency of the data fetching strategy (Section~\ref{subsec:strategy}) depends on data advertisements (Section~\ref{subsec:bitmap}). Let us consider a scenario, where advertisements contain only a few missing packets, while there is a number of missing packets around peers. In this case, the data fetching process may be inefficient, since peers exchange (potentially much) less missing data than what it is actually available around them. To this end, when multiple peers encounter each other for a (potentially) short time, we need to ensure that they quickly become aware of as much available (missing) data as possible around them. To achieve that, we need mechanisms to: (i) prioritize data advertisements from peers that maximize the amount of available (missing) data that encountered peers are aware of, and (ii) mitigate transmission collisions during this process, while at the same time preserve the semantics of data advertisement prioritization.

\noindent \textbf{Data advertisement transmission prioritization:} For the transmission of the first bitmap during an encounter, the peer that has most of the data receives priority. This is useful when a peer having a few (or no) data encounters a peer that has most (or all) of the packets, so that the latter disseminates as much data as possible to the former. For each subsequent transmission, our prioritization strategy\footnote{The prioritization scheme applies only to the transmission of bitmaps, since the RPF strategy determines the order to retrieve data. Collisions during the transmission of Interest and data packets determined by RPF are mitigated through the use of a random transmission timer by each peer.}  is based on the number of packets each peer has that are missing from all the previously transmitted bitmaps. 

\noindent \textbf{Collision mitigation:} Peers can prioritize their bitmap transmissions linearly by dividing a default transmission window by the percent of the packets they have that are missing from previously transmitted bitmaps. This, however, results in frequent collisions when peers have a similar number of packets that are missing from previous bitmaps. To mitigate that, we propose a variant of the Ethernet exponential backoff algorithm, which we call ``Priority-based Exponential Backoff Algorithm (PEBA)''. PEBA separates peers into groups of transmission slots\footnote{The length of the transmission slots can be based on a variety of factors. In the context of this work, we have so far considered the average size of transmitted packets and the channel state (e.g., bandwidth, loss rate).} created through the exponential backoff algorithm, prioritizing peers that have a larger number of missing packets from all the previously transmitted bitmaps. The priority groups and the number of transmission slots are created on a per-encounter basis. 

\begin{figure}[h]
  \centering
  \vspace{-0.3cm}
  \includegraphics[width=8cm]{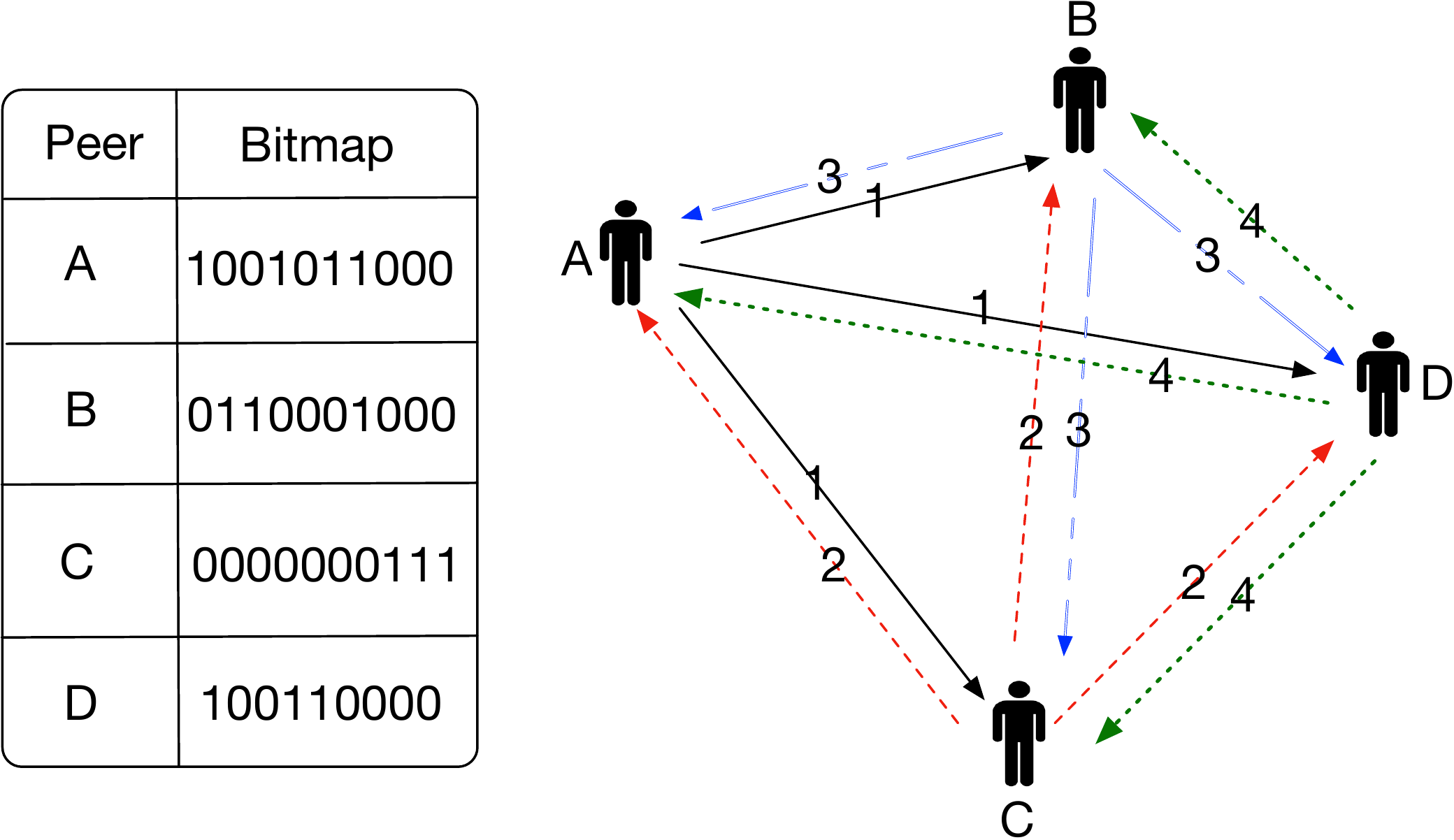}
  \caption{Bitmap prioritization \& collision mitigation example}
  \label{Figure:bitmap-prioritization}
  \vspace{-0.4cm}
\end{figure}

\noindent \textbf{Example:} In Figure~\ref{Figure:bitmap-prioritization}, we assume that peers have a default transmission window and no collisions have occurred. When no collisions have been detected, peers prioritize their bitmap transmissions by dividing the transmission window by the percent of the packets they have that are missing from previously transmitted bitmaps. Therefore, given that A has most of the data, it schedules its transmission timer to expire before others. When peers receive A's bitmap, they cancel their current transmission and reset their timer based on how many packets they have that were missing from A's bitmap. C's timer expires before others, however, B's timer expires before hearing C's transmission. B and C collide and once they detect the collision, PEBA creates two slots. 

We assume that the slots for the peers that collide are divided into two priority groups. Peers that have, at least, half of the missing packets randomly select a slot in the first group, while peers that have fewer than half of the missing packets randomly select a slot in the second group. In our example, there are six packets missing from A's bitmap. C has three, B has two, and D has one of them. Thus, C will be in the first group, while B and D will be in the second group. C transmits during the first slot, while B and D transmit during the second slot, colliding with each other. In this case, PEBA creates four slots for B and D. There are three packets missing from A's and C's bitmaps, therefore B will be in the first group, transmitting during the first or second slot, and D in the second group, transmitting during the third or fourth slot. 

\noindent \textbf{Analysis:} Let us assume that there are $L$ transmission slots in total and peers are divided into $k$ priority groups, thus there are $n = \floor{\frac{L}{k}}$ slots per group. Peers in the $jth$ group select a random slot $s$, where $j*n \leq s < (j+1)*n$. Zhu et al.~\cite{zhu2011performance} proved that the average backoff number before a successful transmission occurs is $N\textsubscript{backoff} = \sum_{i=1}^{\infty} iP\textsubscript{i}$, where $P\textsubscript{i}$ is the probability that a peer has collided $i$ times before a successful transmission. The average delay for a peer to successfully transmit its bitmap is $T\textsubscript{delay} = \frac{L\textsubscript{average} - 1}{2} \tau$, where $L\textsubscript{average} = \frac{n - 1}{2}$ is the peer's average contention window size and $\tau$ the slot duration.

\section {Multi-hop Communication}
\label{sec:extensions}

In this section, we present how DAPES can achieve communication over multiple hops through one or more intermediate nodes that may or may not understand the DAPES semantics. Our multi-hop communication design leverages semantically meaningful naming and the NDN stateful forwarding plane, so that intermediate nodes can make decisions on whether to forward or suppress received Interests based on how likely is that such Interests will retrieve the requested data. 


\subsection {Pure Forwarders}
\label{pure-forwarding}

Peers may meet nodes that do not understand the DAPES semantics (e.g., users that have not installed the application on their device), but understand the NDN network-layer semantics (i.e., only have an NFD instance installed). We call these nodes ``pure forwarders''. Pure forwarders store data transmissions they overhear in their CS, thus satisfying received requests with cached data. They also opportunistically forward Interests based on a probabilistic scheme to: (i) avoid flooding Interests across the network, and (ii) discover data available more than a single hop away. Pure forwarders wait for a random amount of time before forwarding an Interest to: (i) avoid collisions with others, and (ii) avoid unnecessary transmissions, since another node within their communication range might respond to the Interest. 
When they forward an Interest, but do not receive a response, pure forwarders start a suppression timer for the Interest name, not forwarding future Interests with the same name until the timer expires. This timer acts as soft state information that determines whether certain data is currently reachable through a pure forwarder.

In Figure~\ref{Figure:extensions}, the dark (A, D, F, H, K) and grey nodes (C, E, G) are interested in different file collections, while node B is a pure forwarder. We assume that A sends an Interest, which can be a discovery or a bitmap Interest, or an Interest for data. Node B receives this Interest, and further forwards it based on some probability. We assume that B forwards this Interest towards direction 1, without receiving a response, thus it starts a suppression timer for the Interest name.

\subsection {Intermediate Nodes Running DAPES}

Nodes running DAPES store information about the data their neighbors have and the collections their neighbors are interested in. This helps them make adaptive forwarding decisions about the Interests they receive from others. In this way, peers reach others through one or more intermediate peers that are interested in the \emph{same} or a \emph{different} file collection.

\begin{figure}[ht]
	\centering
	\vspace{-0.3cm}
	\includegraphics[width=9cm]{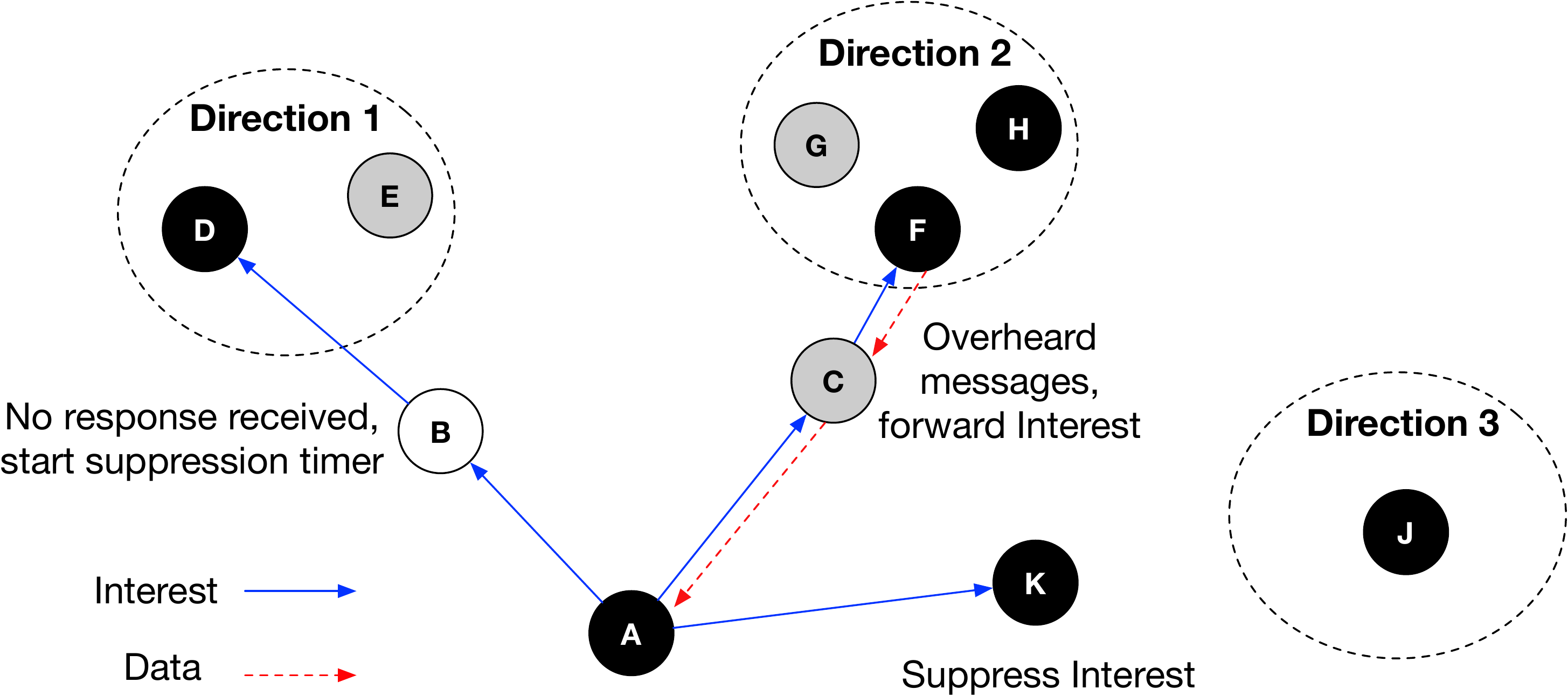}
	\caption{DAPES multi-hop communication example}
	\label{Figure:extensions}
	\vspace{-0.5cm}
\end{figure}

\noindent \textbf{Same file collection:} Intermediate peers interested in the same file collection make forwarding decisions based on their knowledge about the available collection data across their neighbors. In Figure~\ref{Figure:extensions}, K is a direct neighbor of A and both are downloading the same file collection. K knows whether there are peers towards direction 3 that download the same collection and what data they have. Therefore, K forwards received Interests from A only when it speculates that they can bring a response back; for example, when A requests data that K does not have, but J does, or when it is beneficial for A to learn J's bitmap (e.g., J may be able to offer multiple data packets missing from A). 
In our example, we assume that K speculates that forwarding A's Interest to J will not be bring a response back, therefore, K suppresses the Interest.

\noindent \textbf{Different file collections:} Intermediate peers interested in a different file collection make forwarding decisions based on messages they overhear about other collections across their neighbors. If intermediate peers have no knowledge about the requested data (e.g., have not overheard any related messages recently), they follow the probabilistic scheme of pure forwarders, suppressing Interests that do not bring data back. In Figure~\ref{Figure:extensions}, A and F are two hops away, but can reach each other through C, who is interested in a different collection than them. When C receives A's Interest, it decides whether to forward it. For example, if C has overheard the messages between F and H, it knows that F is interested in the same collection as A, thus forwarding A's Interest towards direction 2 will likely retrieve data. Especially if C has overheard F's bitmap, it knows which packets F has, thus being able to accurately decide whether to forward A's Interest. 


\section{Experimental Evaluation}
\label{sec:evaluation}

We performed a simulation study of DAPES to evaluate different design choices (Section~\ref{tradeoffs}) and compare its performance with existing IP-based solutions (Section~\ref{ip-results}). We also performed a DAPES feasibility study through real-world experiments in an outdoor campus setting (Section~\ref{feasibility-study}).

\subsection {Prototype Implementation}

Our DAPES prototype consists of 5K lines of C++ code and uses the ndn-cxx library~\cite{ndn-cxx} to ensure compatibility with NFD. It includes three main components: i) a library that provides the fundamental data structures (e.g., metadata file) and abstractions (e.g., RPF strategy) common for peer-to-peer file sharing in wired and wireless environments, ii) a software module that adapts the library abstractions to our off-the-grid communication environment (e.g., the baseline RPF strategy to provide the different RPF flavors), and iii) an application module that uses abstractions either from the library or the adaptation module to implement the DAPES logic. 

\subsection {Experimental Setup}

We used a collection of image files and experimented with a variable number and size of files. We present the 90th percentile of the results collected after ten trials for simulations and real-world experiments. 

\subsubsection {Simulation Experiments}
\label{simulation-setup}

\begin{figure}[ht]
	\centering
	\vspace{-0.2cm}
	\includegraphics[width=6.5cm]{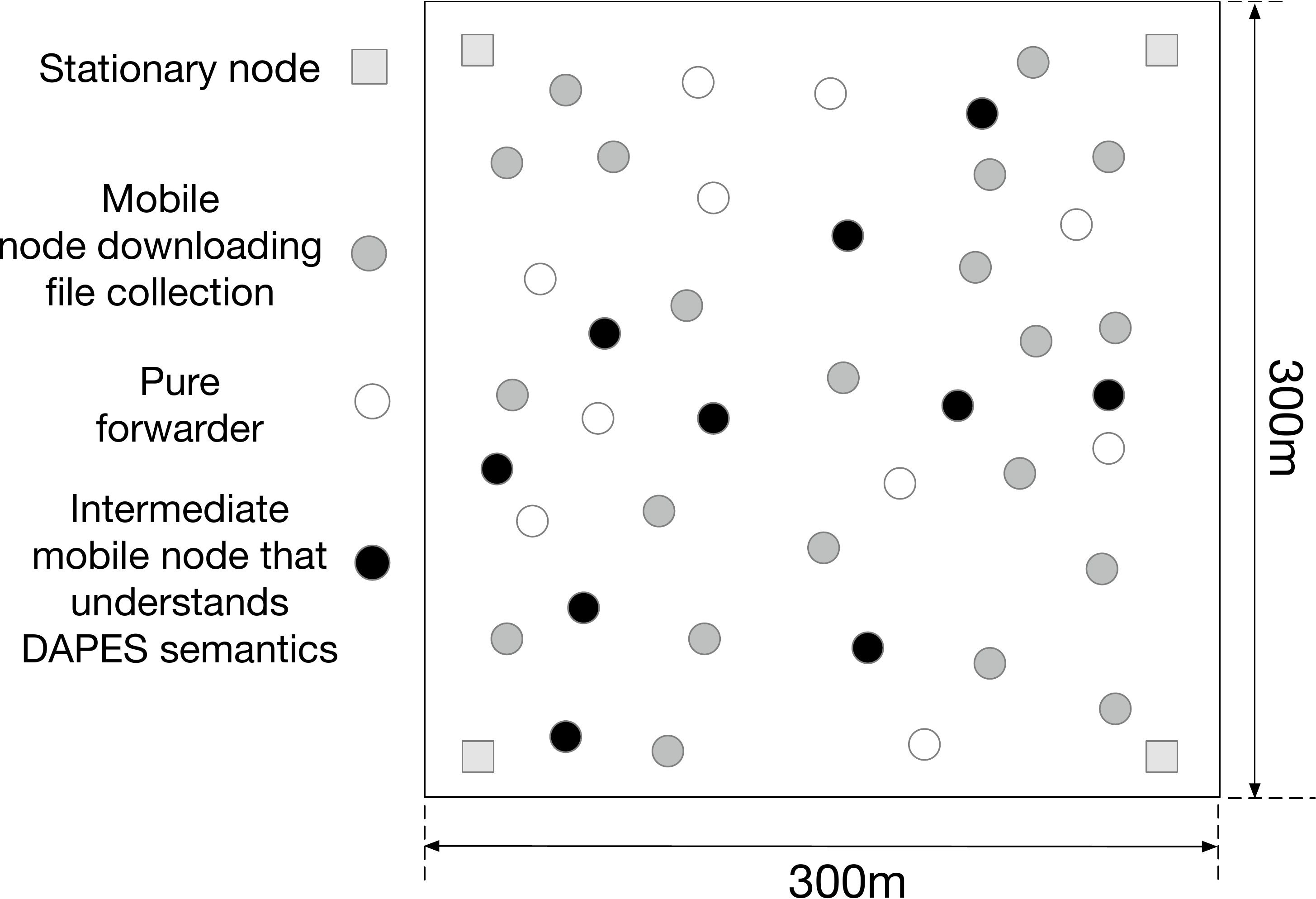}
	\caption{Simulation topology snapshot}
	\label{Figure:simulation-topology}
	\vspace{-0.2cm}
\end{figure}

Our topology (Figure~\ref{Figure:simulation-topology}) consists of 4 stationary (acting as data repositories) and 40 mobile nodes. The mobile nodes randomly choose their direction and speed. The speed ranges from 2m/s to 10m/s and the direction from 0 to 2 $\pi$ (loss rate equal to 10\%). Nodes communicate through IEEE 802.11b 2.4GHz (data rate of 11Mbps). We perform experiments with varying WiFi ranges, which leads to different sizes of connected peer groups over time. The 4 stationary nodes and 20 of the mobiles nodes (randomly chosen) download a \emph{file collection of interest}. Unless otherwise noted, we used a collection of ten files (each file is 1MB and each data packet is 1KB).



\noindent \textbf{DAPES-based experiments:} We ported our DAPES prototype into the ndnSIM simulator~\cite{mastorakis2017evolution}. ndnSIM features software integration with the real-world NDN software prototypes (ndn-cxx and NFD) to offer high fidelity of simulation results. In our topology (Figure~\ref{Figure:simulation-topology}), we randomly select 10 nodes to act as pure forwarders and the remaining 10 nodes understand the DAPES semantics and act as intermediate nodes. Peers use a transmission window of 20ms and select a random value within this window for every transmission other than bitmap transmissions, which are prioritized. Unless otherwise noted, peers use the local neighborhood RPF strategy, interleave data fetching with data advertisements, and fetch advertisements from all the peers within their range. They also communicate over multiple hops (unless otherwise noted the probability of intermediate nodes to forward an Interest is 20\% to ensure message reachability, but also avoid extensive flooding).

\noindent \textbf{IP-based experiments:} We compare DAPES to two IP-based peer-to-peer file sharing solutions for MANET; Bithoc~\cite{krifa2009bithoc, sbai2008bithoc} and Ekta~\cite{pucha2004ekta}. Bithoc peers perform periodic scoped flooding of ``HELLO'' messages to discover others and the data they have. They separate others into ``close'' (at most two hops away) and ``far'' (more than two hops away) neighbors. Peers follow an RPF strategy to fetch data from close neighbors, while they fetch data not available in their nearby neighborhood from far neighbors. Bithoc uses DSDV~\cite{perkins1994highly} as the underlying routing protocol and TCP over IP for reliable delivery. Ekta offers a Distributed Hash Table (DHT) substrate for the search of data objects in MANET by integrating the DHT protocol operations with DSR~\cite{johnson2001dsr} at the network layer. Ekta uses UDP over IP as the transport layer protocol. Following the setup of the DAPES experiments for a fair comparison, in our topology (Figure~\ref{Figure:simulation-topology}), we randomly select 10 nodes to act as forwarders and the last 10 nodes understand the Bithoc and Ekta semantics. All of these 20 nodes forward received packets based on their routing tables.

\noindent \textbf{Evaluation metrics:} We present results (Sections~\ref{tradeoffs} and~\ref{ip-results}) for the following metrics: (i) \emph{file collection download time:} the average time needed for each of the 20 mobile and 4 stationary nodes to download the file collection of interest, and (ii) \emph{transmissions (overhead):} the number of packets transmitted by all the 44 nodes (24 nodes that download the file collection of interest and the 20 intermediate nodes) for the downloading of the collection of interest. For DAPES, the overhead includes the discovery Interests and data, bitmap Interests and data, and the Interest/data packets transmitted for the file collection sharing (including the forwarding transmissions by intermediate nodes). For Bithoc, the overhead includes the packets generated by DSDV, the application-layer flooding, and the TCP overhead for the collection retrieval. For Ekta, the overhead includes the packets generated by DSR for route discovery and maintenance, messages among peers on the DHT to find data packets, and the packets needed to retrieve the file collection.

\label{ip-experiments}

\subsubsection {Real-world Experiments}

We used 5 MacBooks (macOS 10.13), each equipped with an 1.7GHz Intel i7 processor and 8 GB of memory. We ran NDN on top of IEEE 802.11b 2.4GHz (each MacBook had a WiFi range of about 50m). Peers interleaved data fetching with data advertisements and fetched advertisements from all the entities within their range. Peers also used the RPF strategy across their local neighborhood. 

\begin{figure*}
	\centering
	\begin{subfigure}{.32\textwidth}
		\centering
		\includegraphics[scale=0.21]{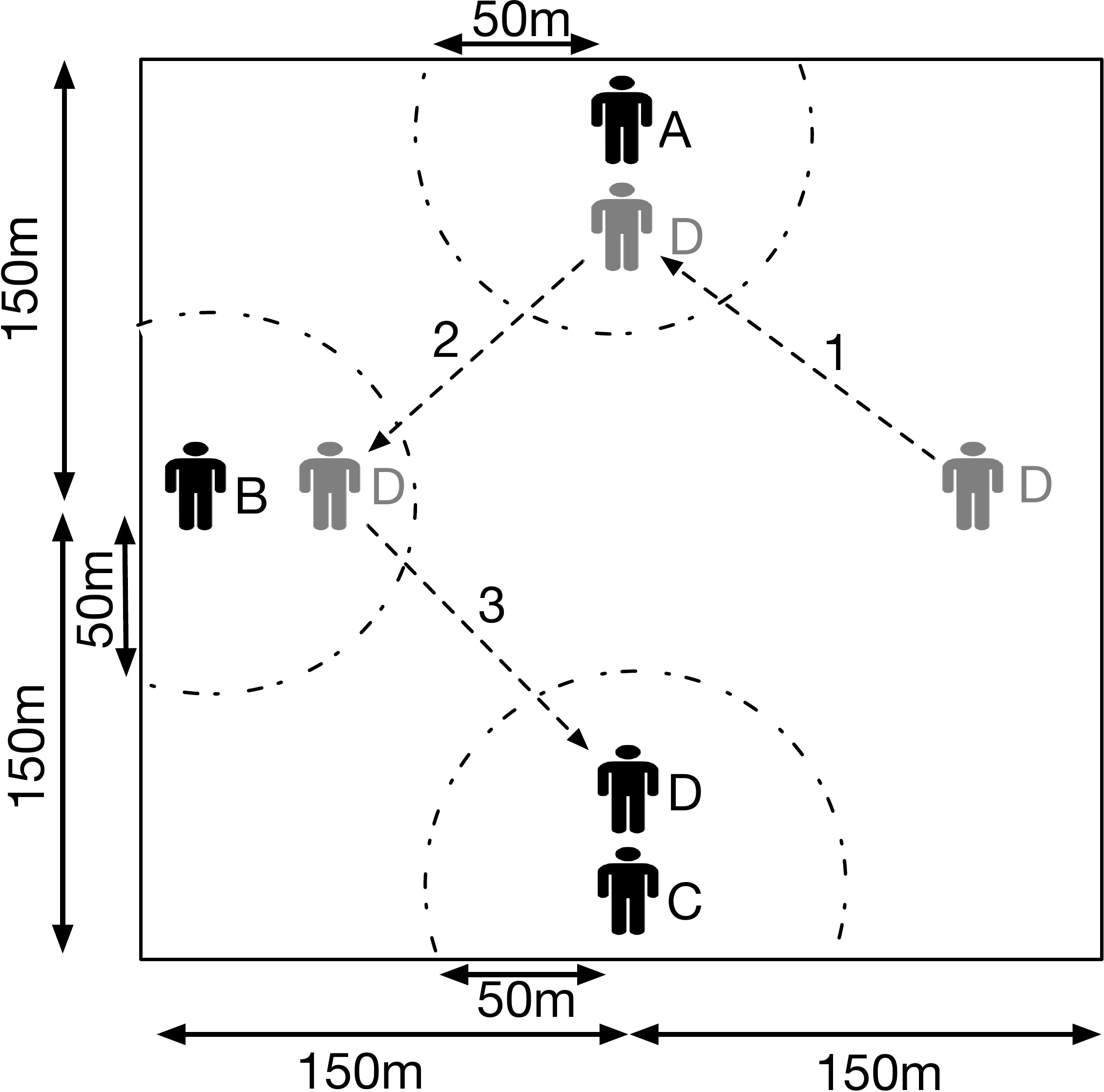}
		\caption{Data sharing through a carrier}
		\label{Figure:scenario1}
	\end{subfigure}
	\begin{subfigure}{.32\textwidth}
		\centering
		\includegraphics[scale=0.215]{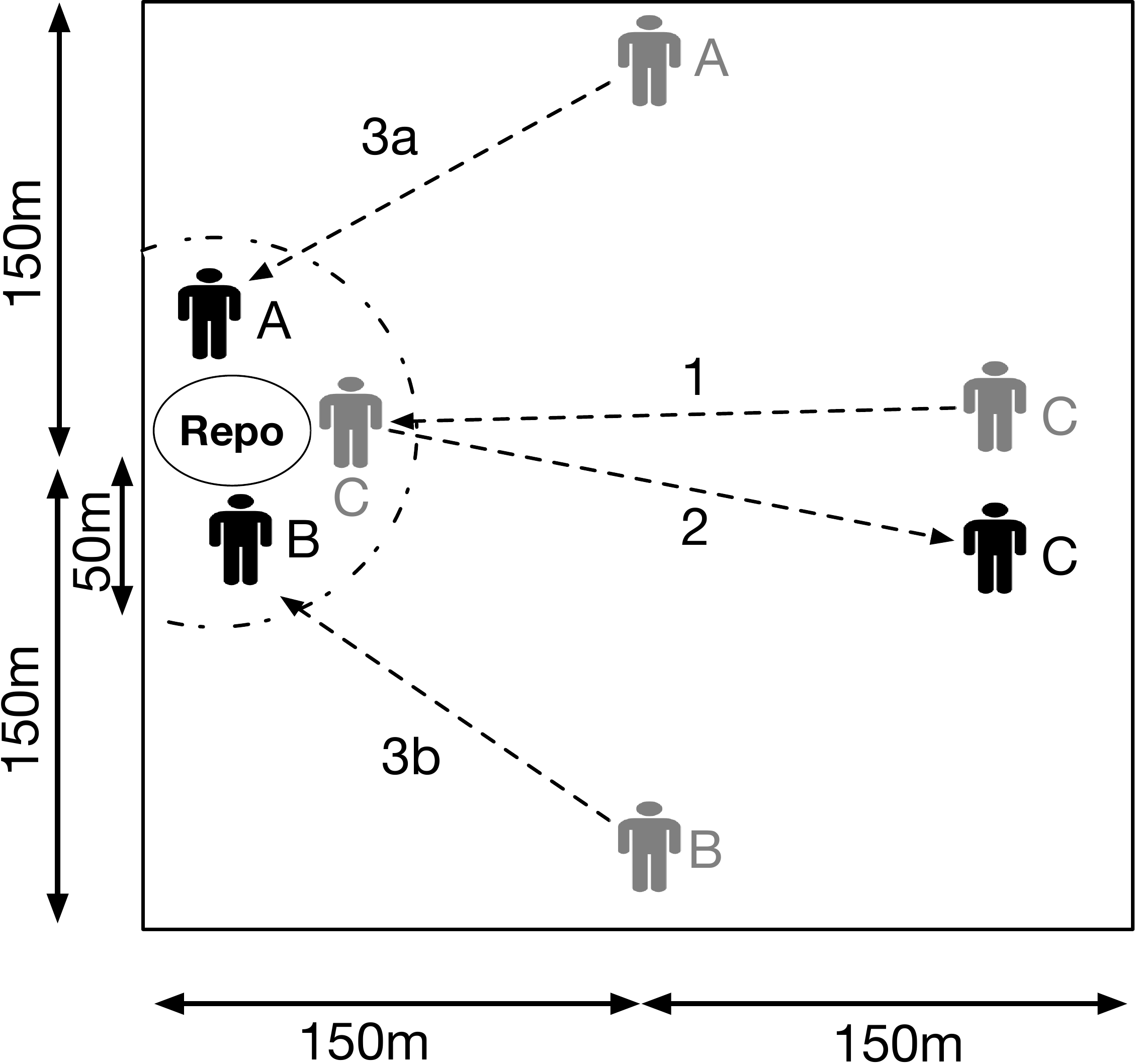}
		\caption{Data sharing through a repository}
		\label{Figure:scenario2}
	\end{subfigure}
	\begin{subfigure}{.32\textwidth}
		\centering
		\includegraphics[scale=0.215]{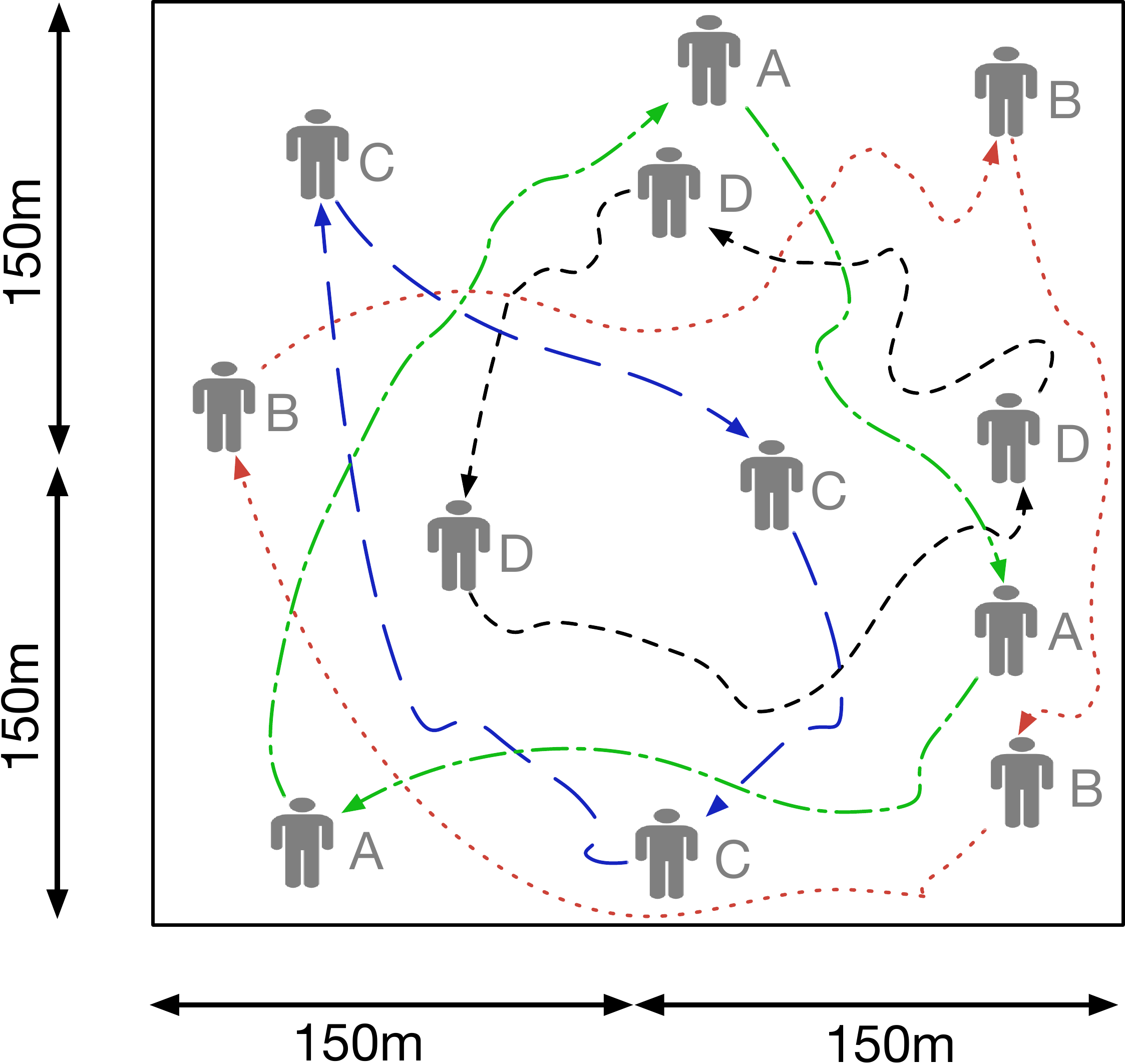}
		\caption{Data sharing among moving nodes}
		\label{Figure:scenario3}
	\end{subfigure}
	\vspace{-0.2cm}
	\caption{Real-world experimental scenarios}
	\label{Figure:topos}
	\vspace{-0.2cm}
\end{figure*}

We experimented with three different scenarios in an outdoor campus setting. In the first one (Figure~\ref{Figure:scenario1}), peer A generates a file collection. D acts as a data carrier that fetches the collection from A and carries it to other network segments, where peers B and C fetch it. In the second one (Figure~\ref{Figure:scenario2}), C generates a collection. The repo downloads the collection from C, while A and B download the collection from the repo. In the third one (Figure~\ref{Figure:scenario3}), A generates a collection that shares with B, C, and D (peers are moving across an area with no infrastructure). To demonstrate how DAPES maximizes the utility of the transmitted data and multi-hop communication, in this scenario, there are moments that all the peers are disconnected and moments that they are within the communication range of each other.

\subsection {DAPES Design Trade-offs}
\label{tradeoffs}

\begin{figure}
	\centering
	\begin{subfigure}[b]{0.49\columnwidth}
		\centering
		\includegraphics[width=\columnwidth]{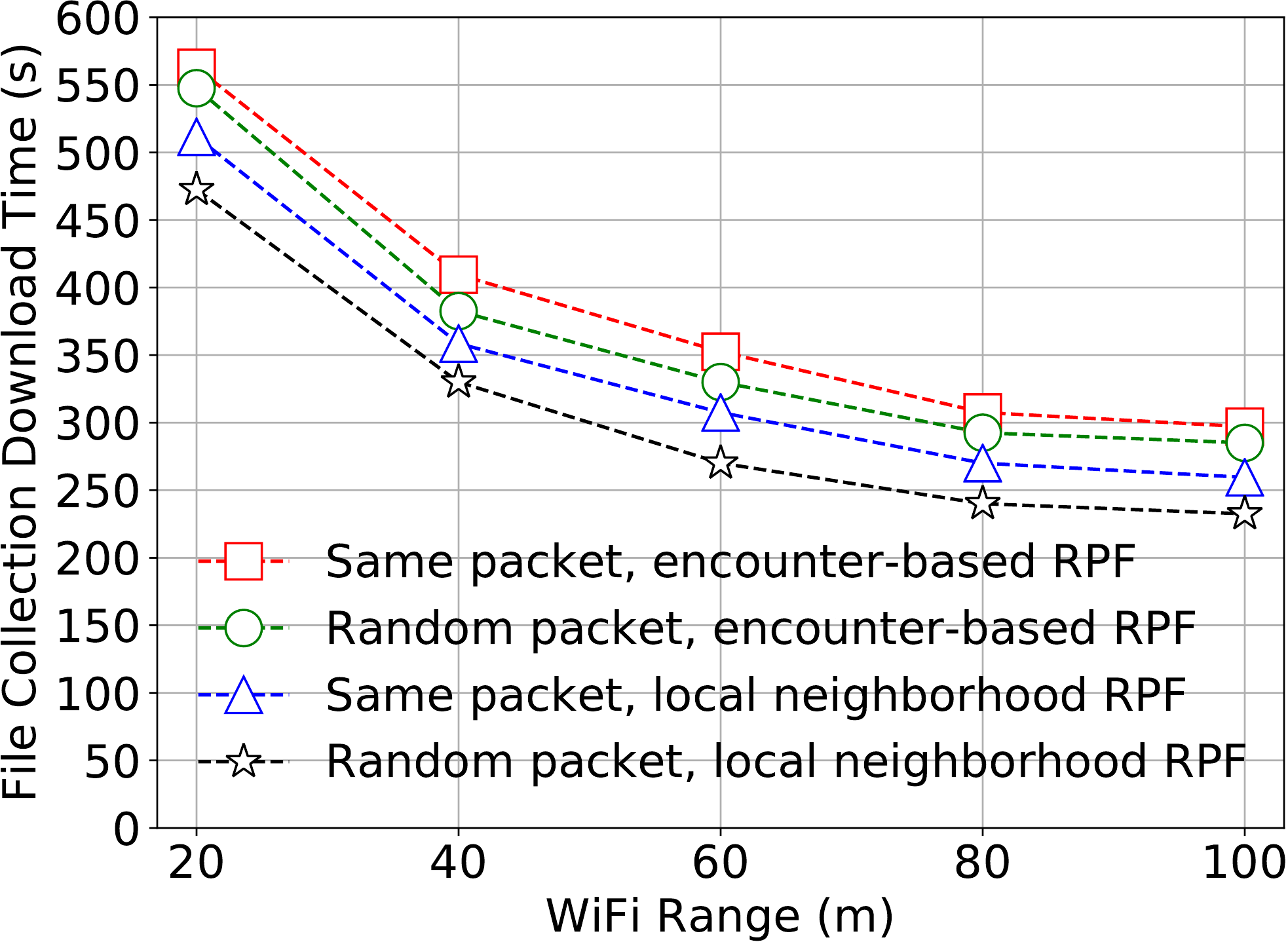}
		\vspace{-0.6cm}
		\caption{File collection download time, different RPF strategies}
		\label{Figure:rpf} \hfil
	\end{subfigure}
	\begin{subfigure}[b]{0.49\columnwidth}
		\centering
		\includegraphics[width=\columnwidth]{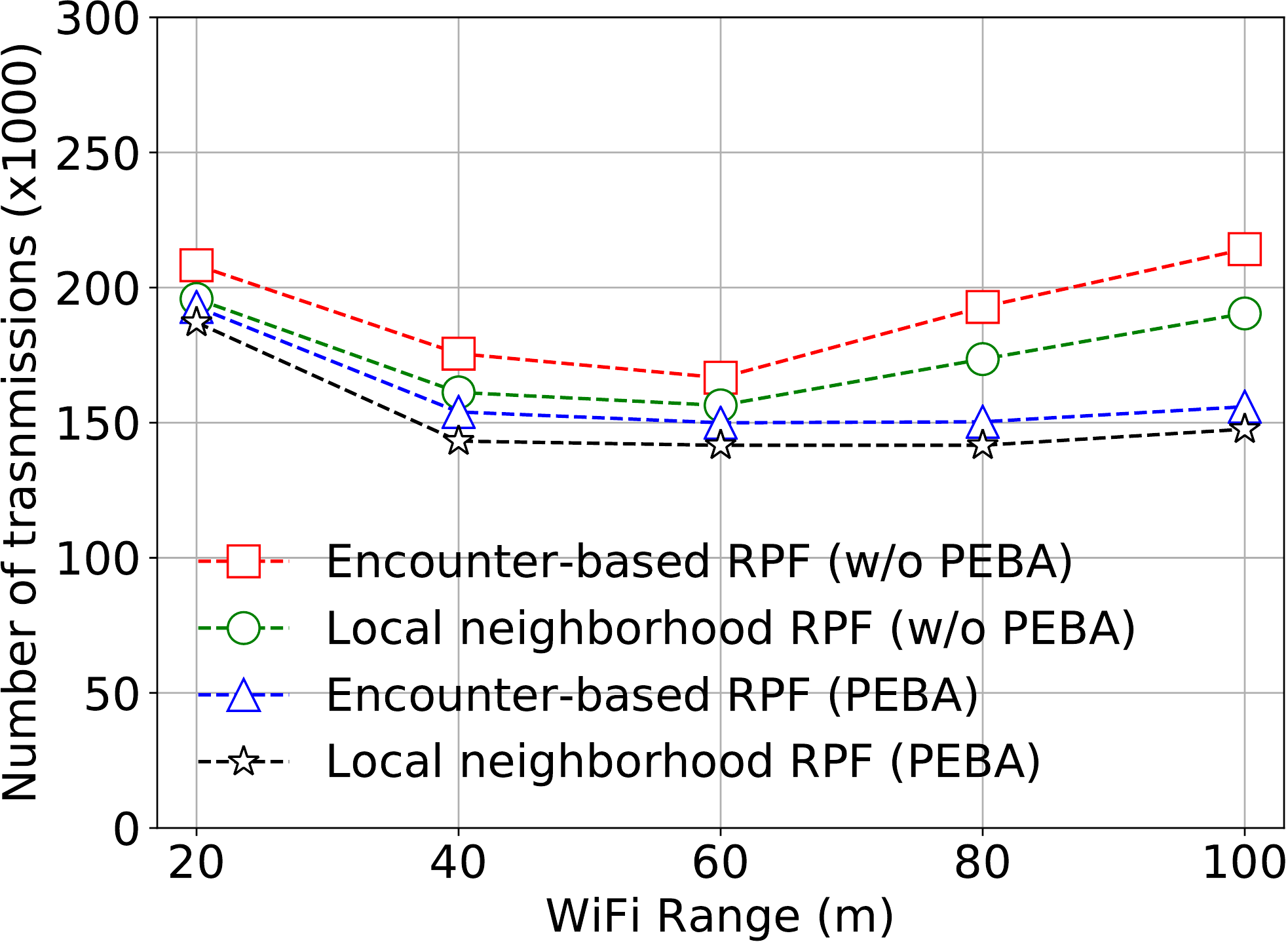}
		\vspace{-0.6cm}
		\caption{Transmissions, different RPF strategies (with and w/o PEBA)}
		\label{Figure:transmissions} \hfil
	\end{subfigure}
	\begin{subfigure}[b]{0.49\columnwidth}
		\centering
		\includegraphics[width=\columnwidth]{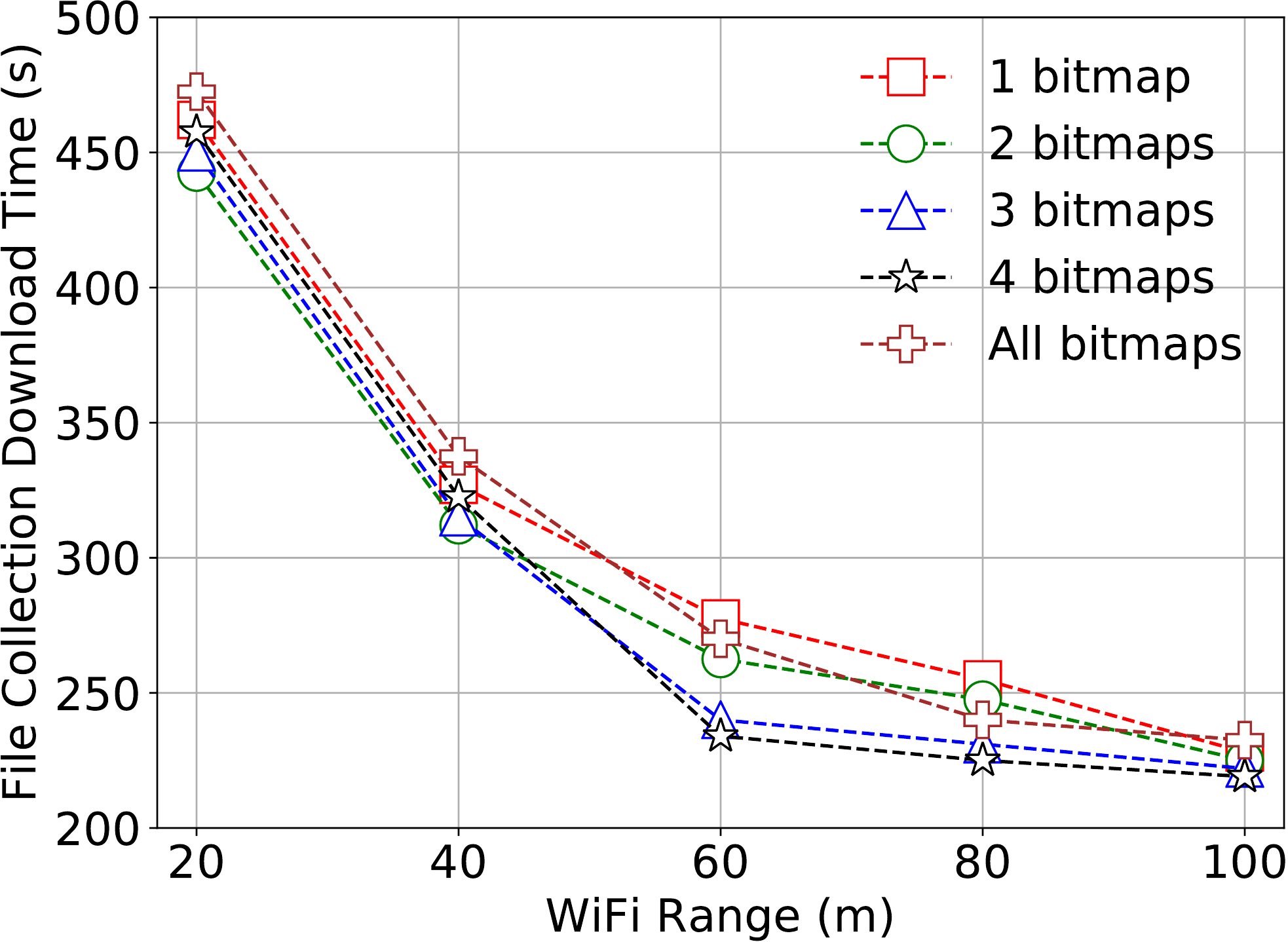}
		\vspace{-0.6cm}
		\caption{File collection download time, bitmap exchanges before data download}
		\label{Figure:bitmaps-before-data} \hfil
	\end{subfigure}
	\begin{subfigure}[b]{0.49\columnwidth}
		\centering
		\includegraphics[width=\columnwidth]{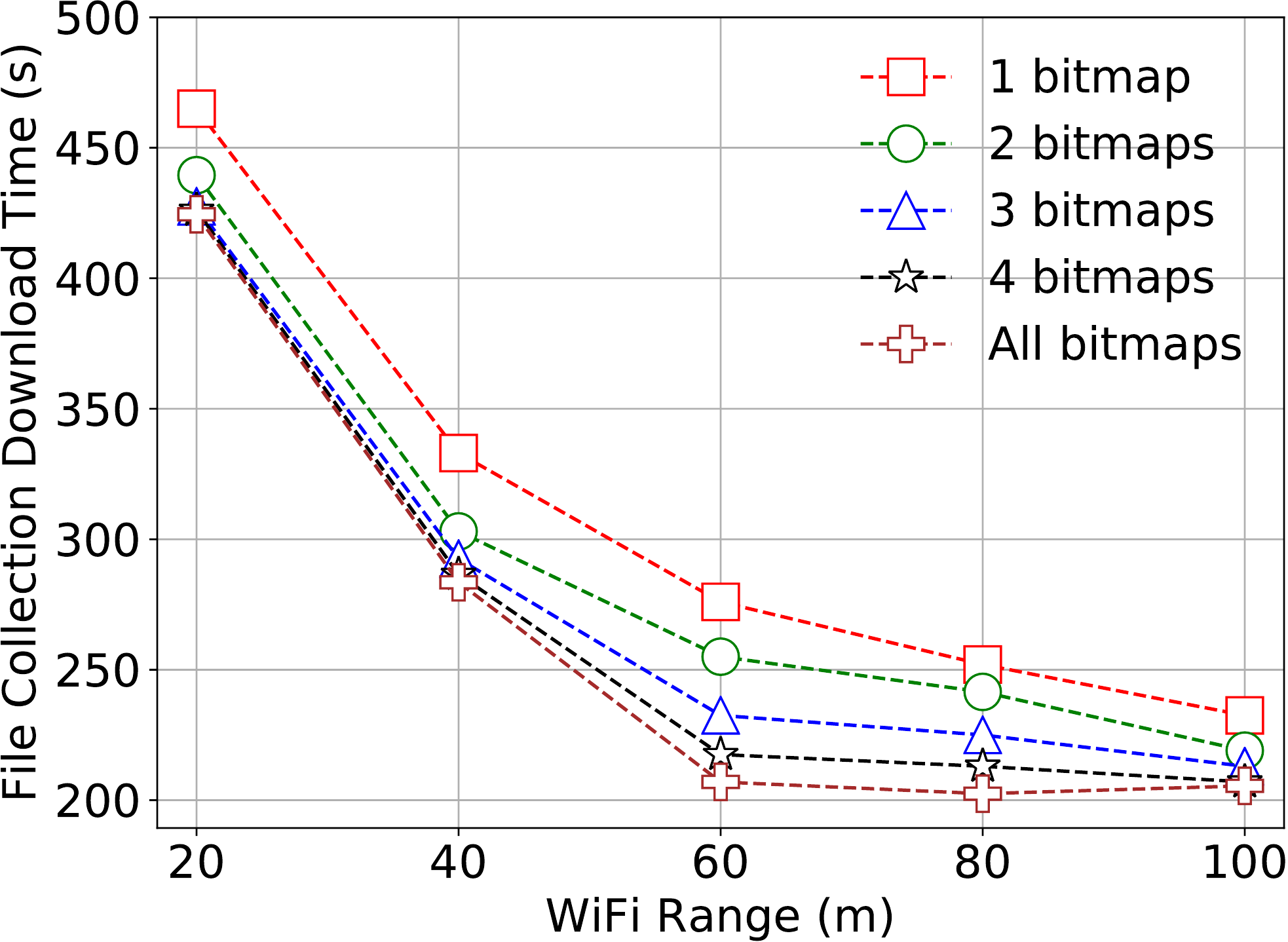}
		\vspace{-0.6cm}
		\caption{File collection download time, bitmap exchanges during data download}
		\label{Figure:bitmaps-during-data} \hfil
	\end{subfigure}
	\begin{subfigure}[b]{0.49\columnwidth}
		\centering
		\includegraphics[width=\columnwidth]{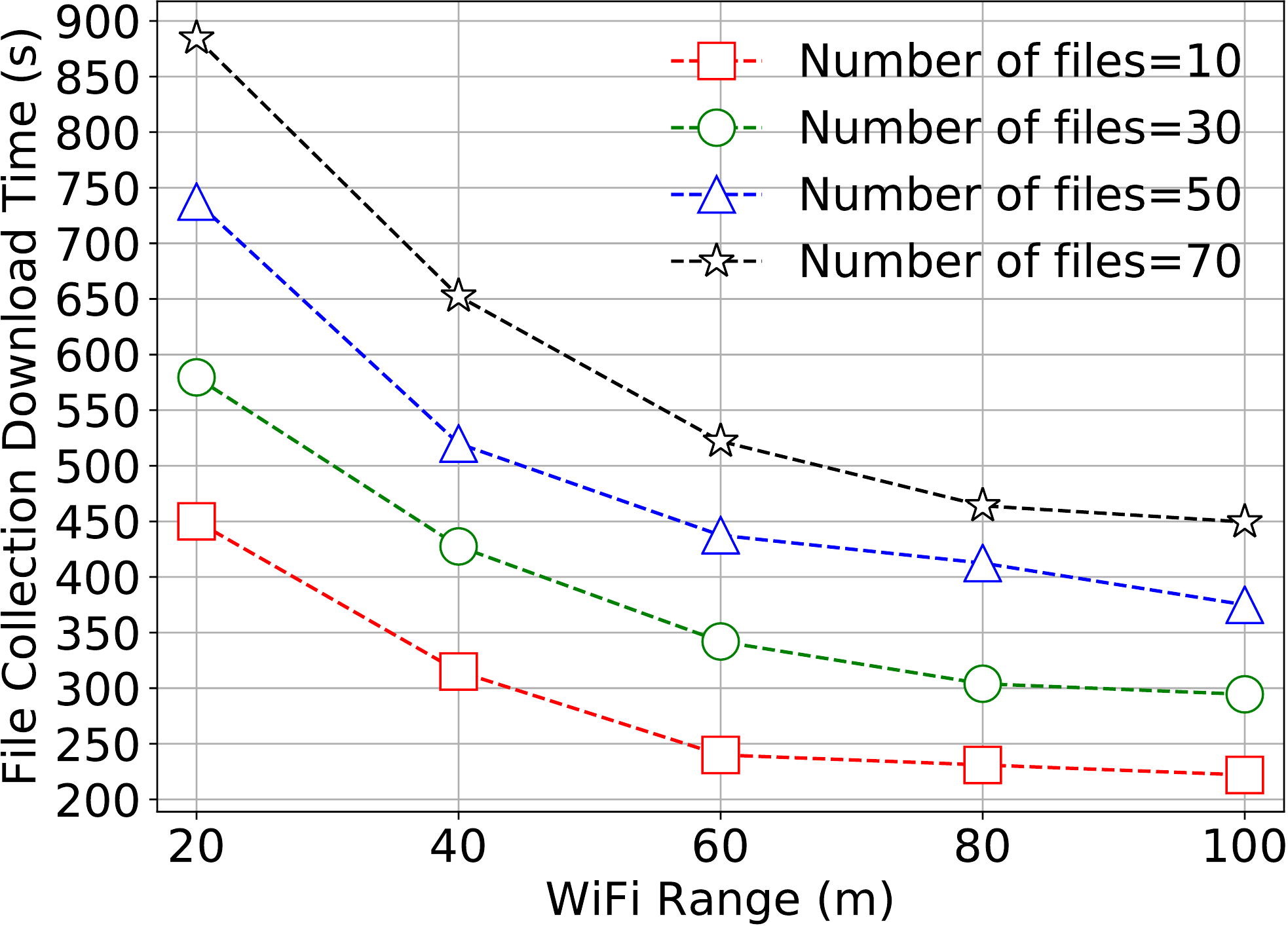}
		\vspace{-0.6cm}
		\caption{File collection download time, varying number of files}
		\label{Figure:file-number} \hfil
	\end{subfigure}
	\begin{subfigure}[b]{0.49\columnwidth}
		\centering
		\includegraphics[width=\columnwidth]{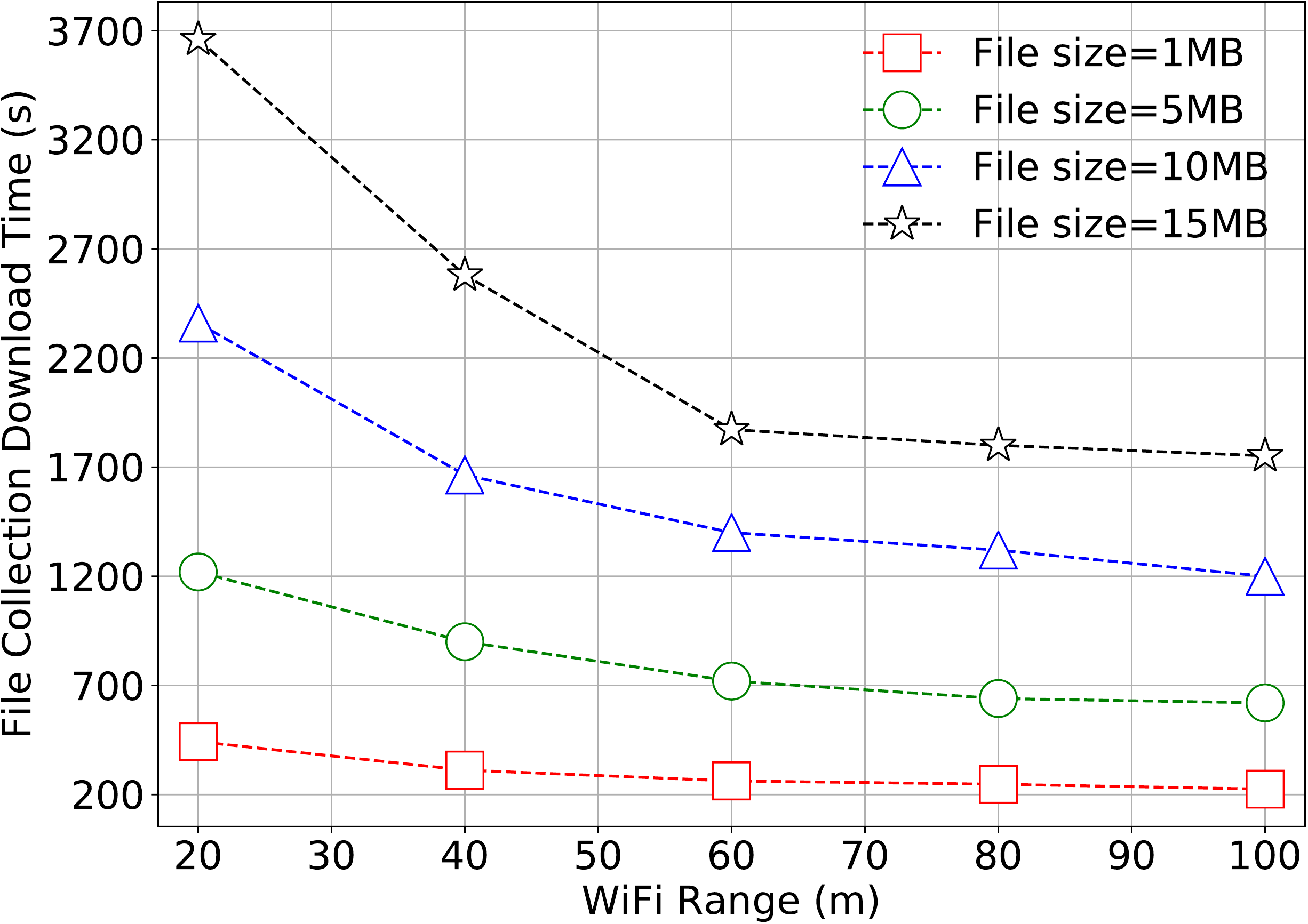}
		\vspace{-0.6cm}
		\caption{File collection download time, varying size of files}
		\label{Figure:file-size} \hfil
	\end{subfigure}
	\begin{subfigure}[b]{0.49\columnwidth}
		\centering
		\includegraphics[width=\columnwidth]{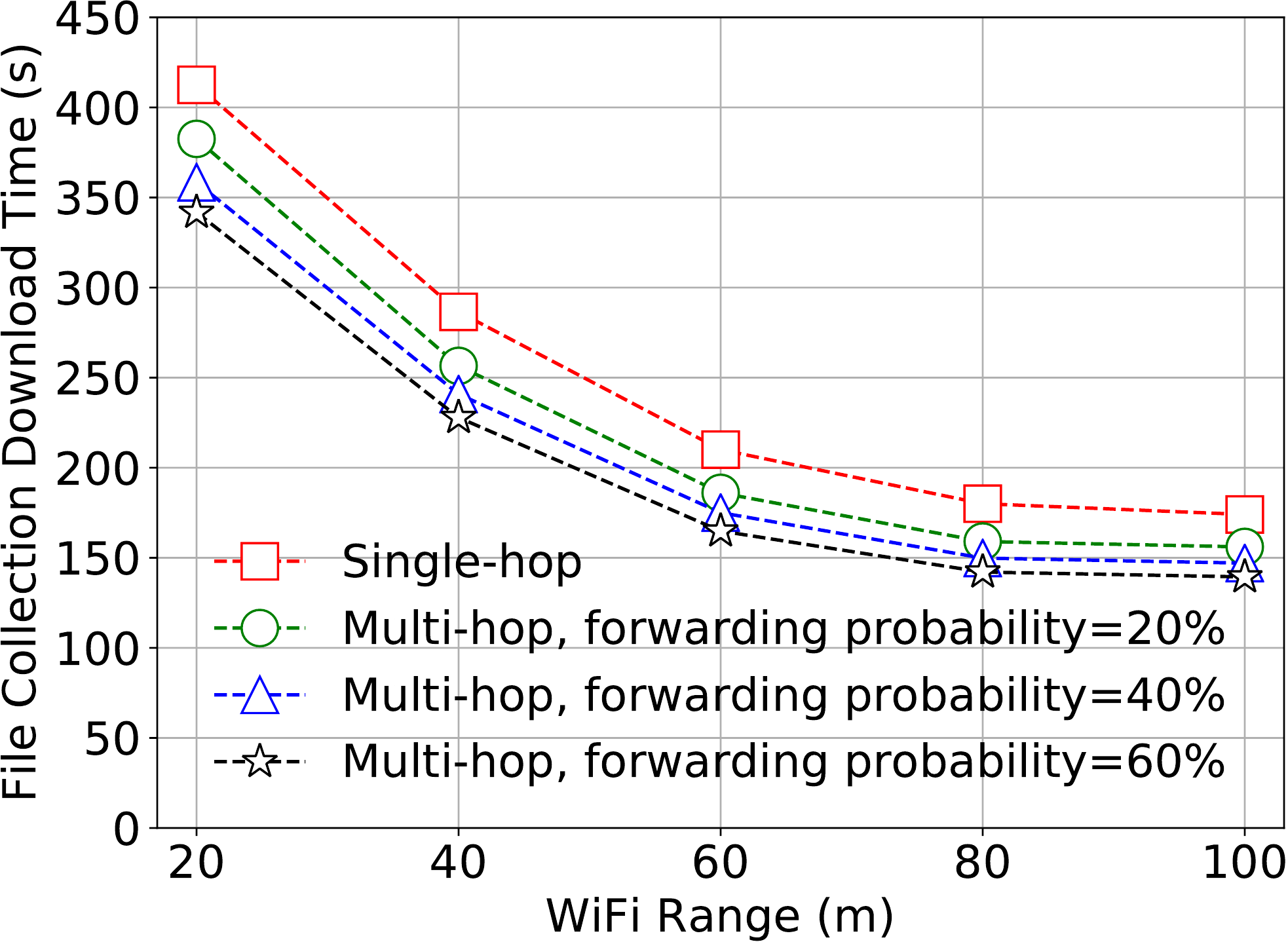}
		\vspace{-0.6cm}
		\caption{File collection download time, varying forwarding probability}
		\label{multihop-delay} \hfil
	\end{subfigure}
	\begin{subfigure}[b]{0.49\columnwidth}
		\centering
		\includegraphics[width=\columnwidth]{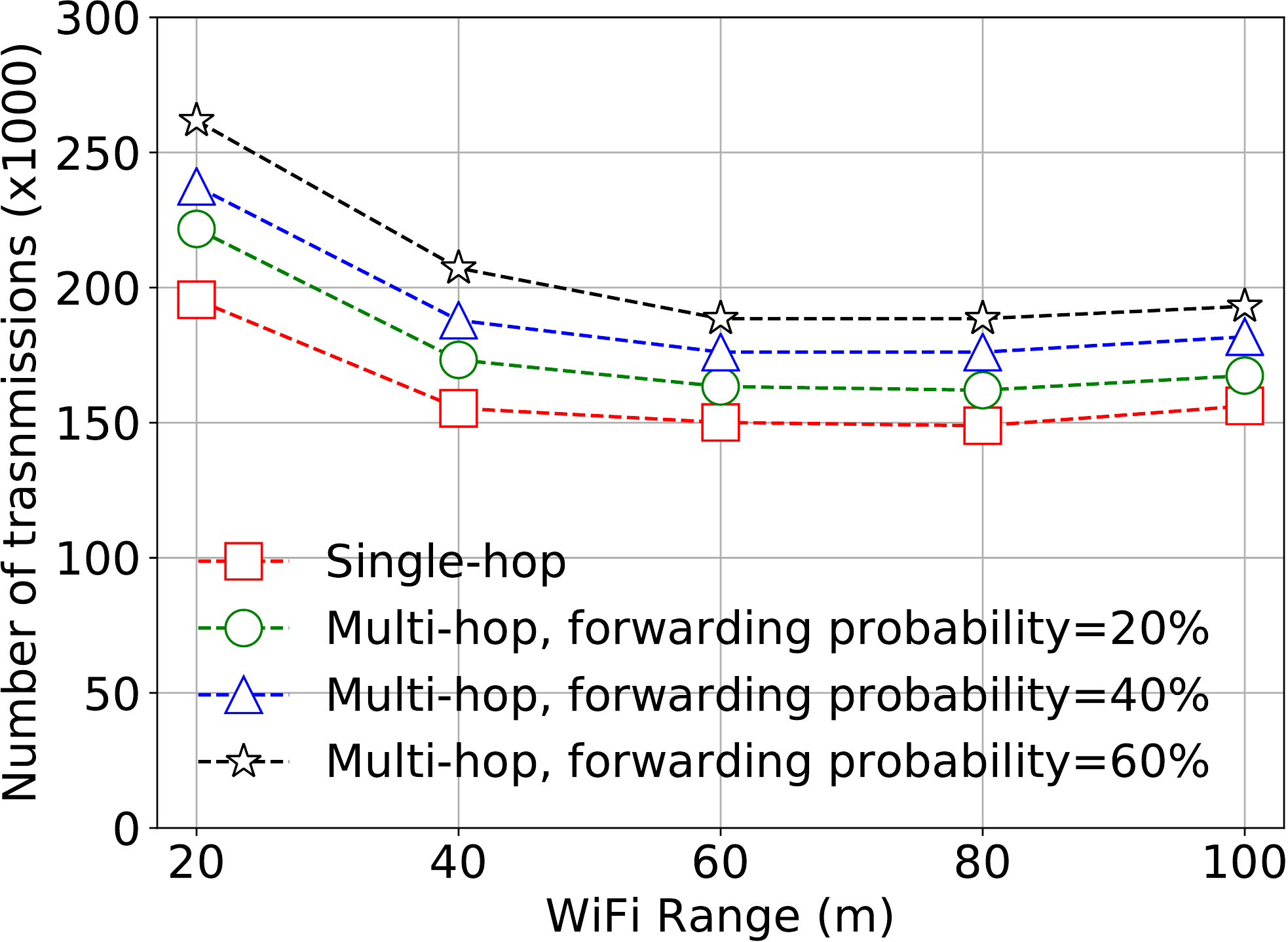}
		\vspace{-0.6cm}
		\caption{Transmissions, varying forwarding probability}
		\label{multihop-overhead} \hfil
	\end{subfigure}
	\label{results}
	\vspace{-0.8cm}
	\caption{DAPES design trade-off results}
\end{figure}


\begin {comment}

\begin{figure*}
	\centering
	\begin{subfigure}{.32\textwidth}
		\centering
		\includegraphics[scale=0.28]{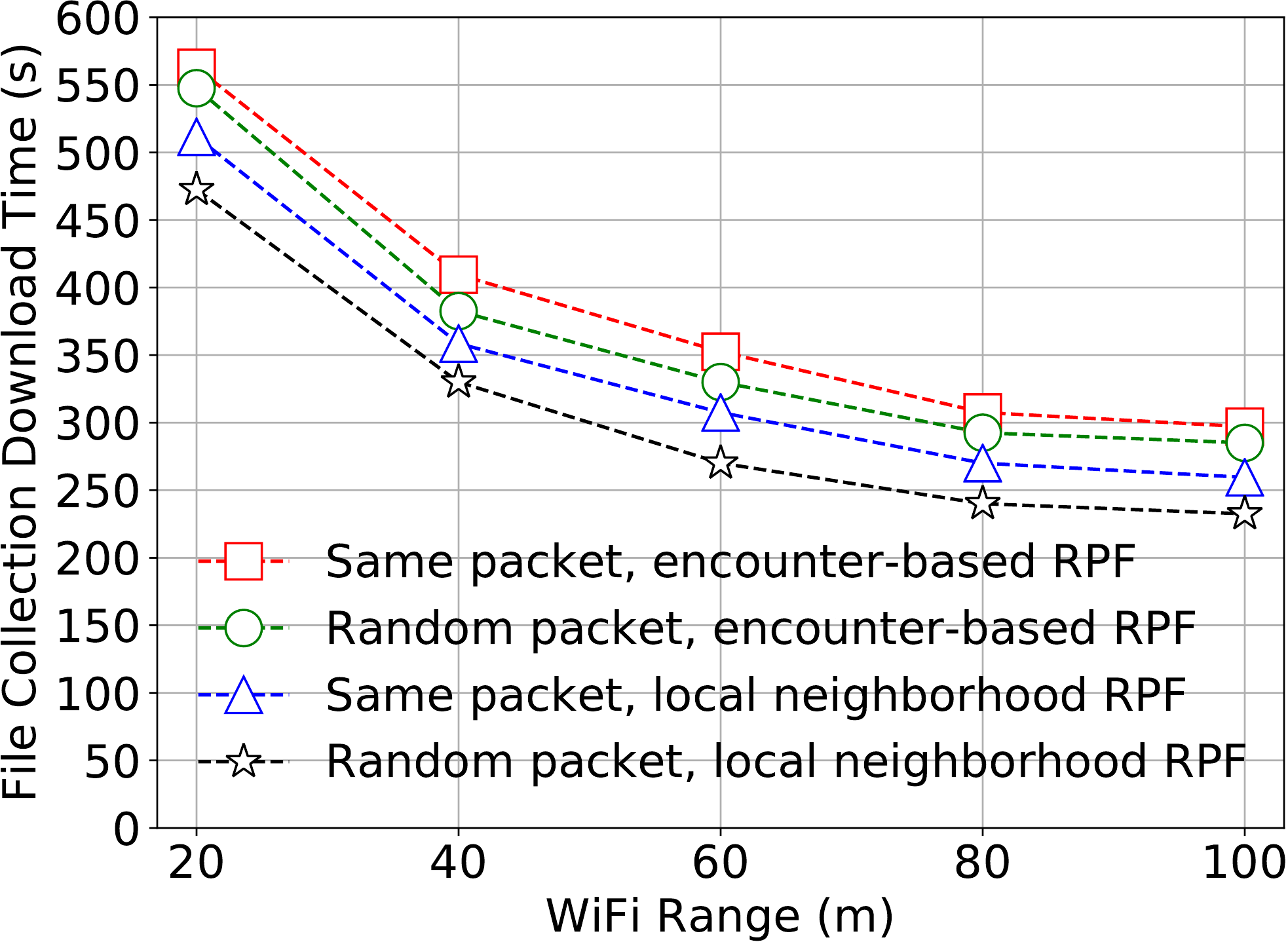}
		\vspace{-0.2cm}
		\caption{File collection download time, different RPF strategies}
		\label{Figure:rpf} \hfil
	\end{subfigure}
	\begin{subfigure}{.32\textwidth}
		\centering
		\includegraphics[scale=0.28]{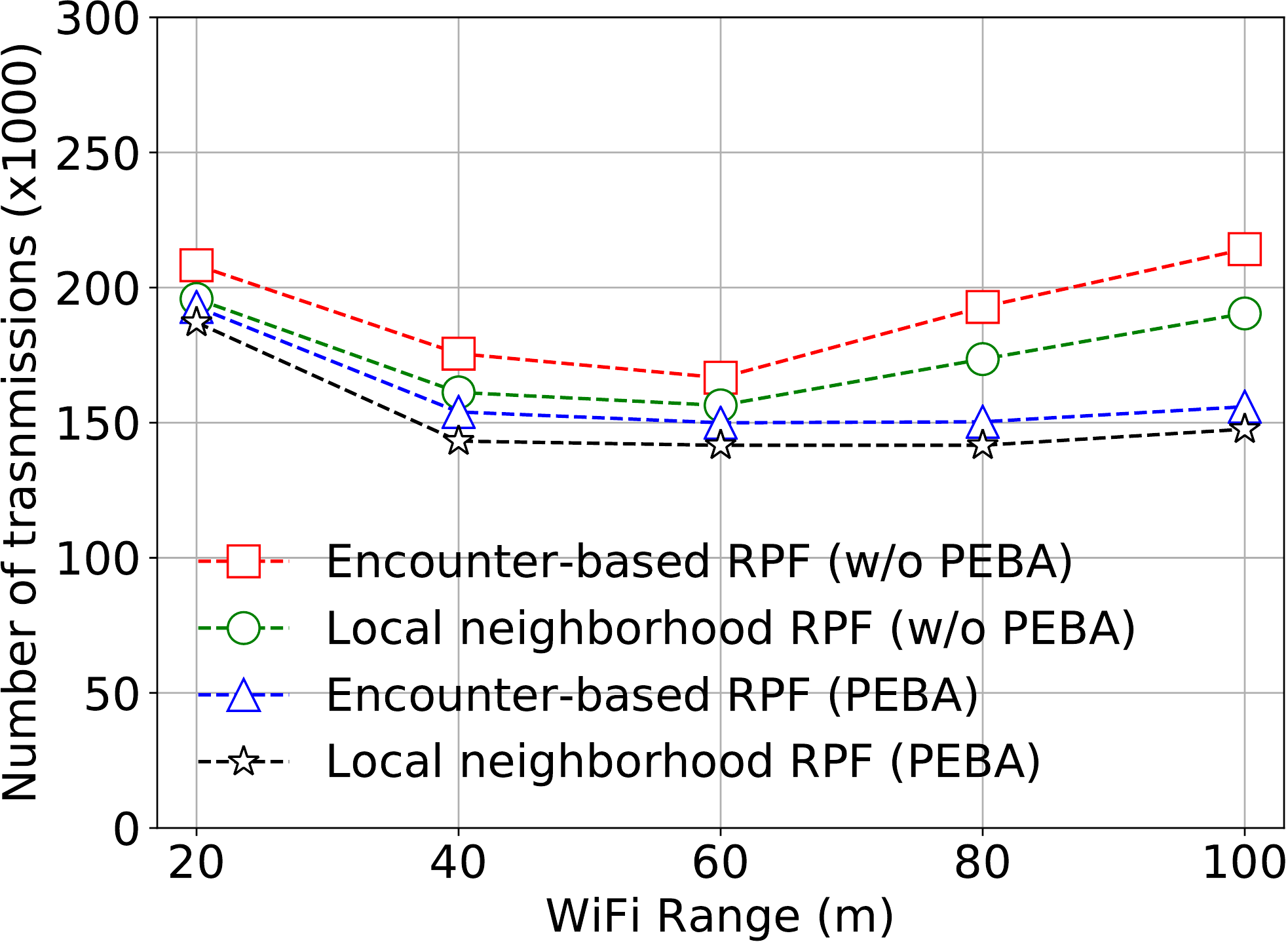}
		\vspace{-0.2cm}
		\caption{Transmissions, different RPF strategies (with and without PEBA)}
		\label{Figure:transmissions} \hfil
	\end{subfigure}
	\begin{subfigure}{.32\textwidth}
		\centering
		\includegraphics[scale=0.28]{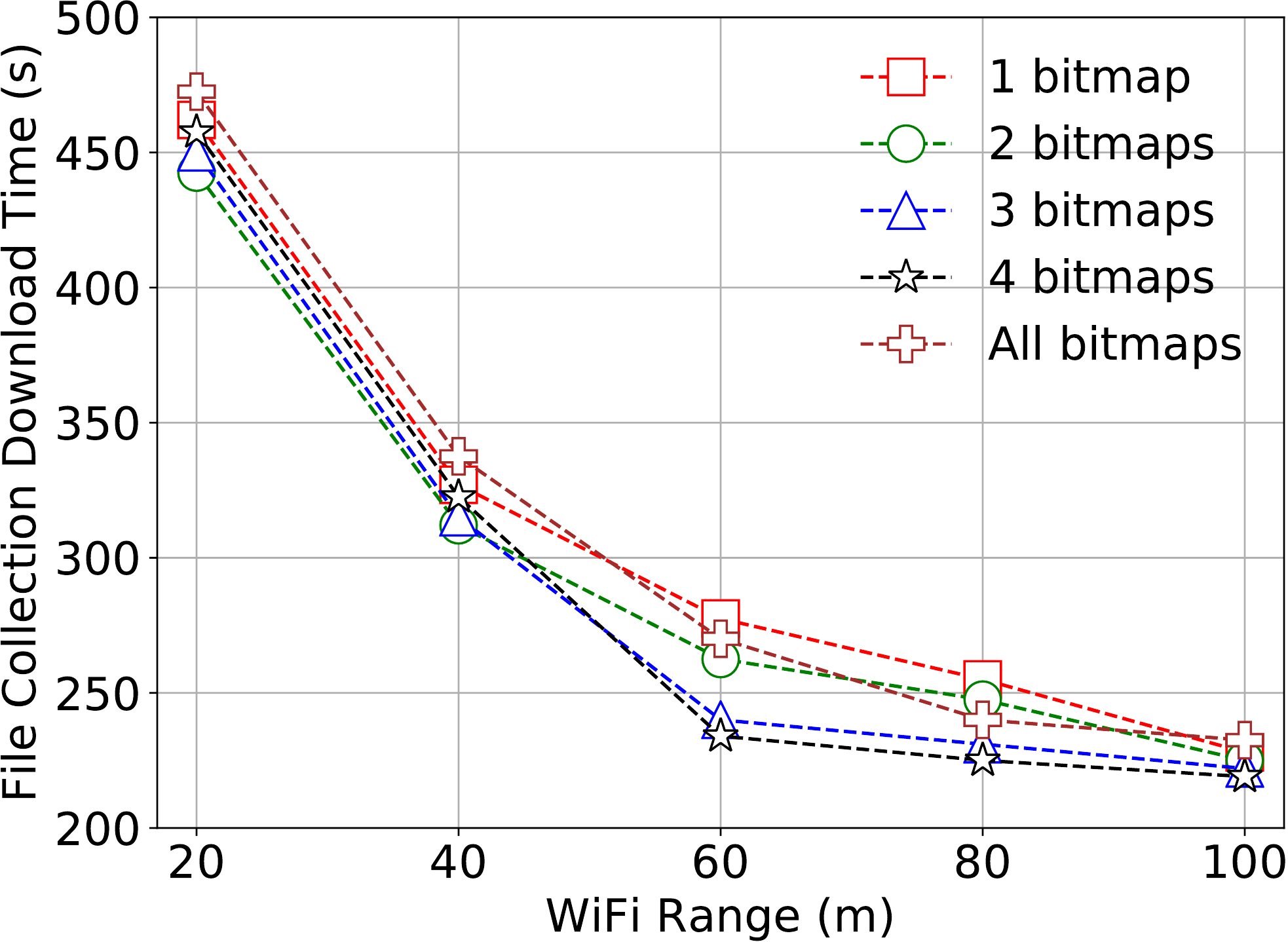}
		\caption{File collection download time, bitmap exchanges before data download}
		\label{Figure:bitmaps-before-data} \hfil
	\end{subfigure}
	\begin{subfigure}[b]{.32\textwidth}
		\centering
		\includegraphics[scale=0.28]{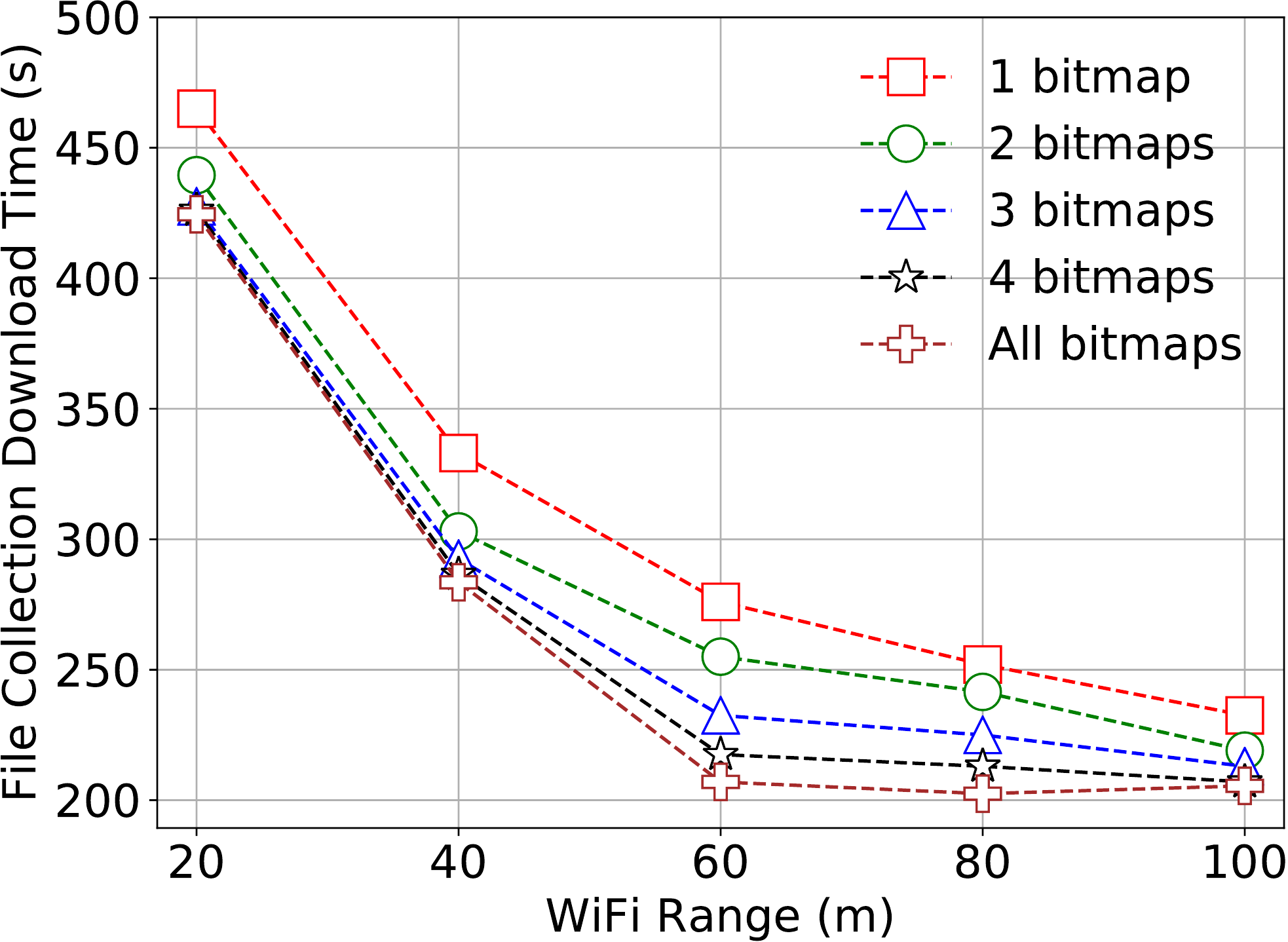}
		\caption{File collection download time, bitmap exchanges during data download}
		\label{Figure:bitmaps-during-data} \hfil
	\end{subfigure}
	\begin{subfigure}[b]{0.32\textwidth}
		\centering
		\includegraphics[scale=0.28]{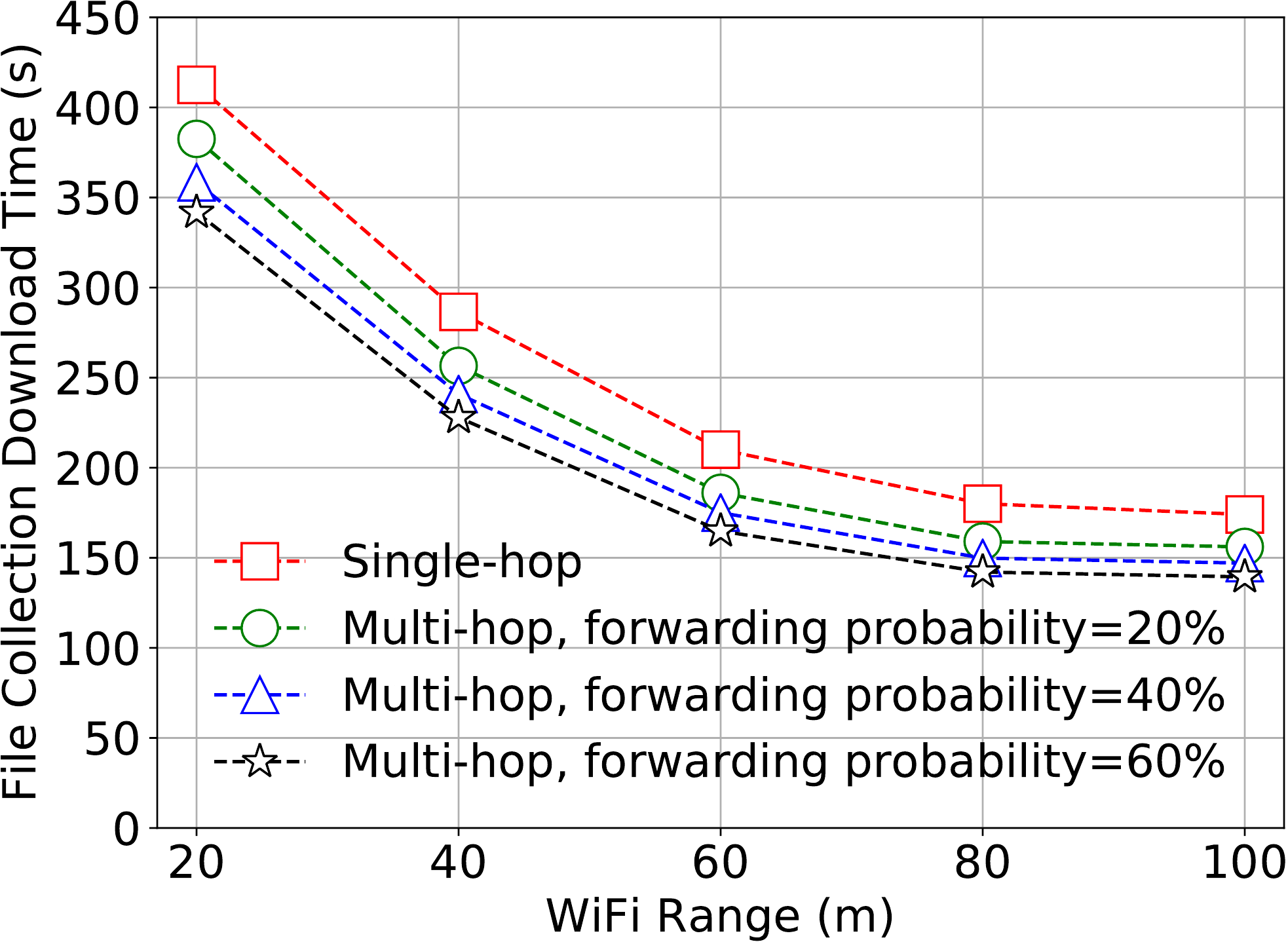}
		\caption{File collection download time, varying forwarding probability}
		\label{multihop-delay} \hfil
	\end{subfigure}
	\begin{subfigure}[b]{.32\textwidth}
		\centering
		\includegraphics[scale=0.28]{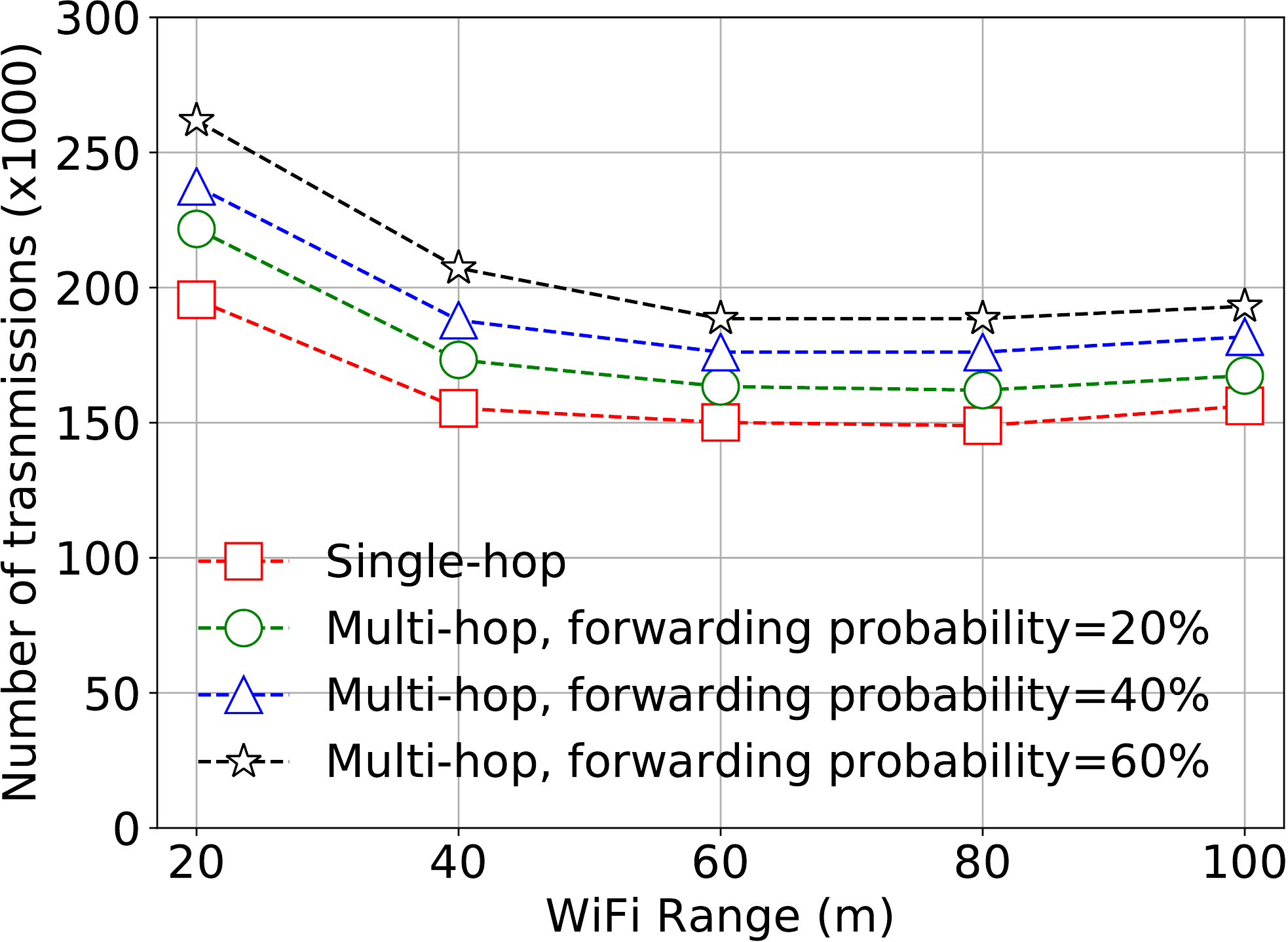}
		\caption{Transmissions, varying forwarding probability}
		\label{multihop-overhead} \hfil
	\end{subfigure}
	\label{results}
	\vspace{-0.5cm}
	\caption{DAPES design trade-off results}
	\vspace{-0.6cm}
\end{figure*}

\end {comment}

\noindent \textbf {Data fetching strategy:} In Figure~\ref{Figure:rpf}, we present the download time for the encounter-based and local neighborhood RPF strategies when peers first fetch the bitmap of all the others within their communication range and then share data. The results show that the local neighborhood strategy performs about 12-14\% better than the encounter-based. The former strategy focuses on retrieving the data missing by most of the peers within their current neighborhood, while the latter also considers previous encounters among peers that might not be within the communication range of each other anymore. As a result, fewer transmissions take place among peers when the local neighborhood strategy is used (Figure~\ref{Figure:transmissions}). 

The results also show that when peers start their downloading process with a random rather than the same packet of the file collection, they are able to download the collection about 11-15\% faster. Starting with a random packet in the file collection helps peers retrieve different blocks of data, thus increasing the diversity of the disseminated data. Note that as we increase the WiFi range (more peers are directly connected to each other), the download time decreases at a slower rate. We conclude that this is due to collisions, given that the number of transmissions for both strategies increases with the WiFi range as shown in Figure~\ref{Figure:transmissions}. 

\noindent \textbf {Collision mitigation:} Figure~\ref{Figure:transmissions} shows the number of transmissions for both flavors of the RPF strategy with and without PEBA (Section~\ref{subsec:prioritization}). Without PEBA, both strategies result in a large number of transmissions as the WiFi range increases. This is due to collisions for the bitmap transmissions; to prioritize bitmap transmissions, peers divide their transmission window by the percent of packets they have, which are missing from previously transmitted bitmaps. As the WiFi range increases, more peers are directly connected. This results in more peers that have similar data and, as a consequence, schedule their transmissions very close to each other. On the other hand, PEBA reduces the number of transmissions by 22-28\%, since it mitigates bitmap transmission collisions through an exponential backoff mechanism, which at the same time preserves the bitmap prioritization semantics.


\noindent \textbf {Data advertisement exchange strategy:} In Figure~\ref{Figure:bitmaps-before-data}, we present the download time when peers exchange their bitmaps first and then download data for a varying number of exchanged bitmaps. The results show that peers minimize the download time when they have enough knowledge about the available data around them, so that the RPF strategy can make effective decisions (2-3 bitmaps for shorter and 4 bitmaps for longer WiFi ranges respectively). Peers move away from each other if they spend too much time exchanging bitmaps before downloading data (e.g., in the illustrated ``all bitmaps" case, where peers fetch the bitmap of every other peer within their communication range). This conclusion verifies our analysis in Section~\ref{subsec:bitmap}. 

In Figure~\ref{Figure:bitmaps-during-data}, we present the download time when peers interleave their bitmap and data exchanges. The results demonstrate the benefit from fetching data as soon as peers have any knowledge about the available data around them. As they collect more bitmaps, the RPF strategy becomes more effective and peers download data faster. This interleaved bitmap and data fetching strategy results in 16-23\% shorter download times than the strategy of fetching bitmaps first and then data.

\noindent \textbf {Variable number and size of files:} In Figures~\ref{Figure:file-number} and~\ref{Figure:file-size}, we present the download time for a varying number of files per collection (each file is 1MB) and varying sizes of collection files (each collection has ten files) respectively. As expected, the download time increases with the total amount of data to be shared. The results demonstrate that the properties of DAPES hold as the collection size grows.



\noindent \textbf{Impact of intermediate nodes:} In Figure~\ref{multihop-delay}, we present the download time for a varying forwarding probability by intermediate nodes, while Figure~\ref{multihop-overhead} shows the number of packets transmitted for the file collection retrieval. When pure forwarders and intermediate DAPES nodes with no knowledge about the requested data forward 20-60\% of received Interests, the collection download time decreases by 12-23\% compared to the results for the DAPES single-hop design. On the other hand, the overhead (packet transmissions) increases by 14-38\%. Overall, the results show that it is adequate for pure forwarders and intermediate DAPES nodes with no knowledge about the requested data to forward 20-40\% of the received Interest. Forwarding a larger amount of Interests results in little performance gain and substantial overhead increase. 

	


\subsection {Comparison to IP-based Solutions}
\label{ip-results}

\begin{figure}
	\centering
	\begin{subfigure}[b]{0.49\columnwidth}
		\centering
		\includegraphics[width=\columnwidth]{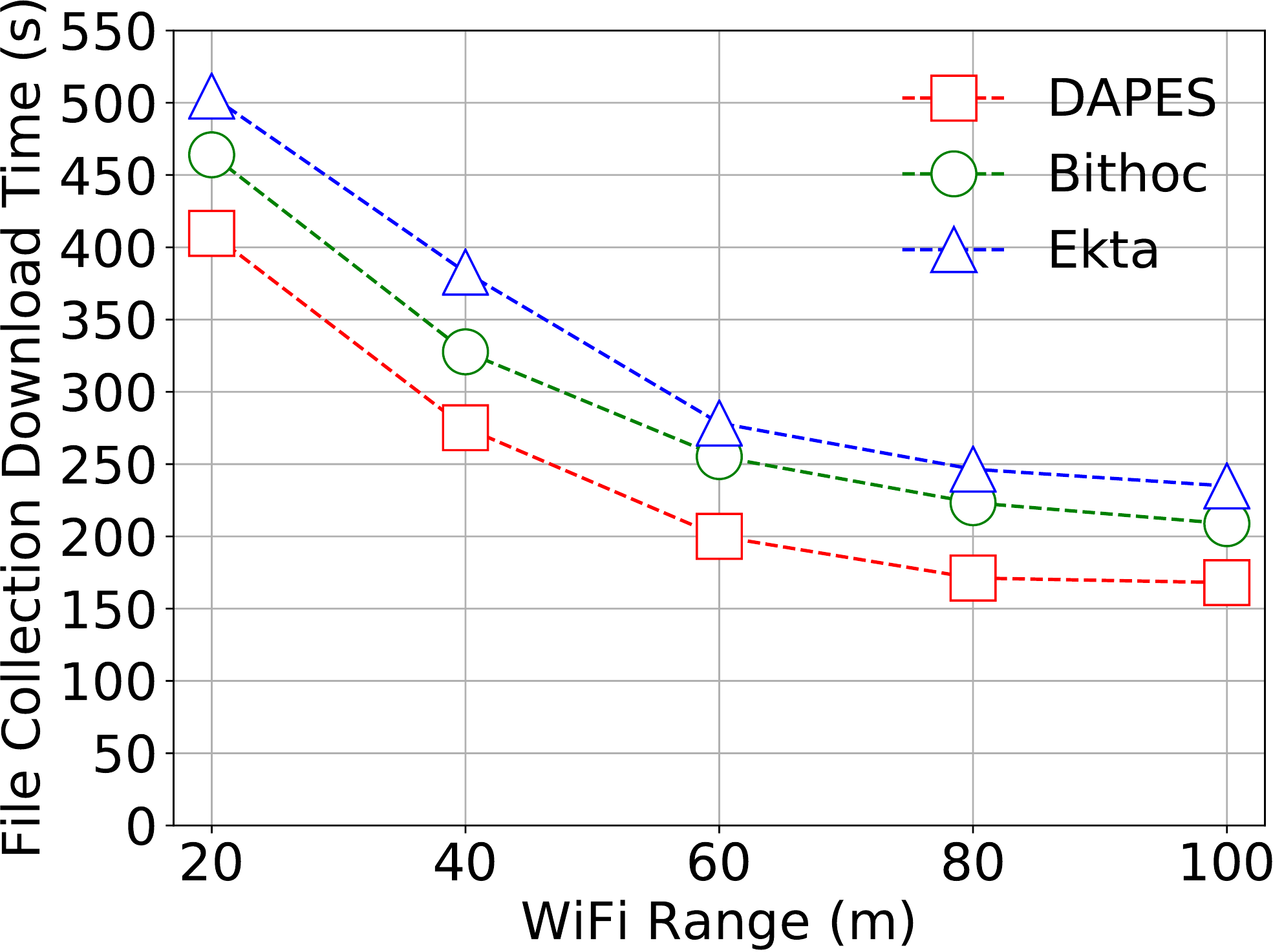}
		\vspace{-0.3cm}
		\vspace{-0.3cm}
		\caption{Download time}
		\label{comparison-delay}
	\end{subfigure}
	\begin{subfigure}[b]{.49\columnwidth}
		\centering
		\includegraphics[width=\columnwidth]{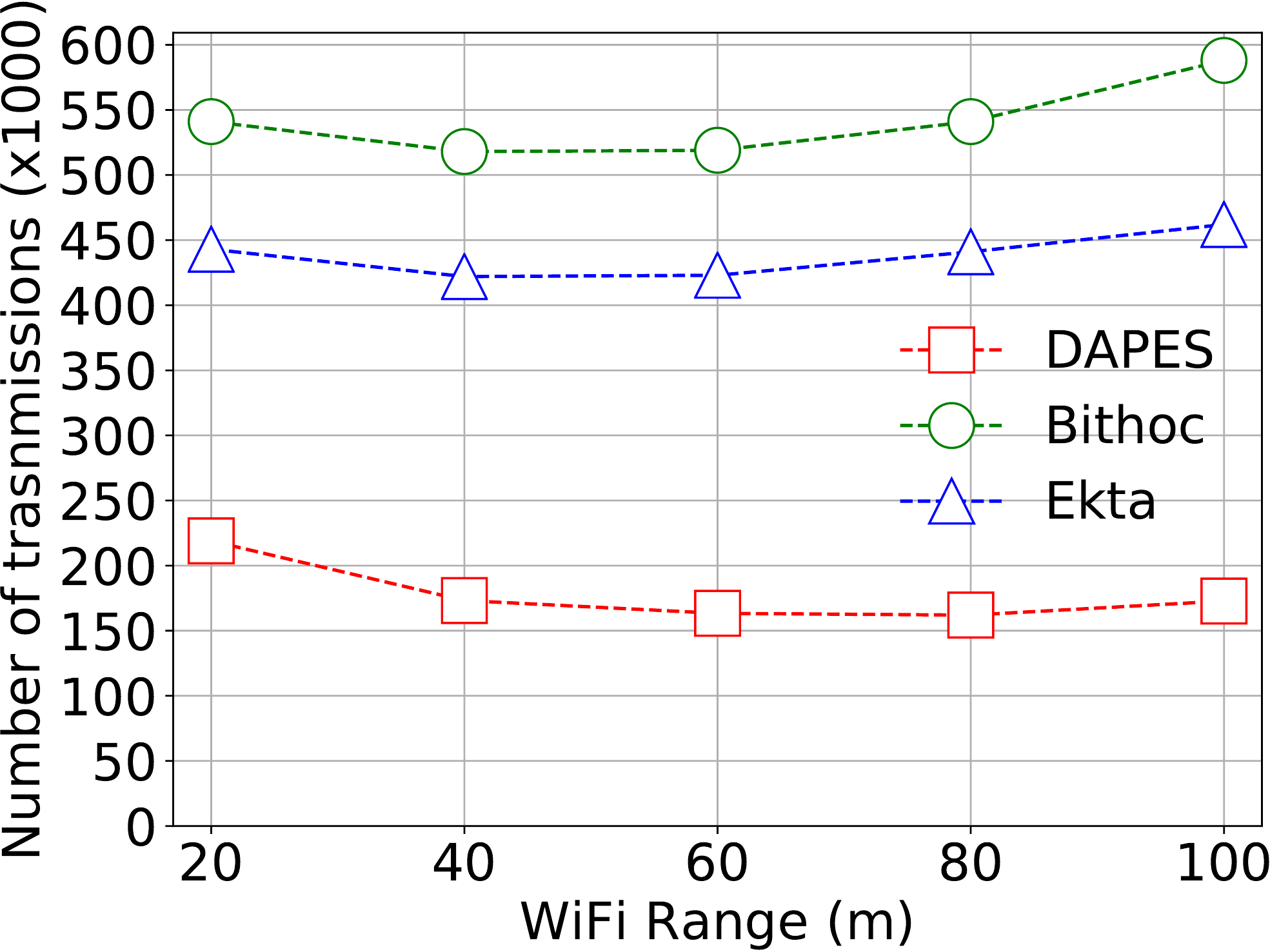}
		\vspace{-0.3cm}
		\vspace{-0.3cm}
		\caption{Transmissions}
		\label{comparison-overhead}
	\end{subfigure}
	\label{tcpip-results}
	\vspace{-0.5cm}
	\caption{Comparison to IP-based solutions}
	\vspace{-0.7cm}
\end{figure}

\noindent\textbf{File collection download time:} In Figure~\ref{comparison-delay}, we present the download time results, which show that DAPES achieves 15-27\% and 19-33\% lower download times than Bithoc and Ekta respectively. Bithoc and Ekta identify individual receivers (based on their IP address) for each packet they send. Even if multiple peers missing common data are within the communication range of the sender, a separate packet has to be sent to each one of them. Paths to each peer need to be established first, then discover what data a peer has, and then begin data retrieval. On the other hand, the semantically meaningful naming of DAPES enables peers to identify the data missing by most of the peers around them, maximizing the utility of transmissions. 
DAPES decouples data sharing from the location of nodes within the network; data requests can be satisfied with data close to the requester, while intermediate nodes can satisfy Interests with cached data.

\noindent\textbf{Transmissions:} Figure~\ref{comparison-overhead} presents the number of transmissions for each solution. DAPES results in 62-71\% and 50-59\% lower overheads than Bithoc and Ekta respectively. Bithoc relies on proactive routing to maintain routes towards peers, and application-layer messages to discover what data each peer has. Due to intermittent connectivity, established routes break and the TCP performance degrades over multiple wireless hops~\cite{tcpmanet}. Ekta is based on reactive routing, resulting in lower overheads than Bithoc (routes are maintained on-demand). In DAPES, each data packet is useful to multiple peers, while intermediate nodes forward or suppress received Interests based on the data available around them. This results in accurate forwarding decisions (83\% of the forwarded Interests successfully brought data back) and low overheads.

\vspace{-0.2cm}

\subsection {Real-World Feasibility Study}
\label{feasibility-study}

In Table~\ref{Table:exp-results}, we present the results for each scenario of Figure~\ref{Figure:topos}. Overall, these results verify the conclusions of our simulation study. 
DAPES can offer with a single transmission data needed by multiple peers, while multi-hop communication comes with its own cost in terms of system load, since peers need to store information about the data available around them.

In the first scenario (Figure~\ref{Figure:scenario1}), the communication involves only two parties (the data carrier and a peer in each connected group), therefore, more time and transmissions are needed for all the entities to download the file collection. In the second scenario (Figure~\ref{Figure:scenario2}), peers A and B fetch the file collection from the repository at the same time. Data requested by either A or B can satisfy both, therefore, the collection can be downloaded faster with fewer transmissions. In the third scenario (Figure~\ref{Figure:scenario3}), peers take advantage of the time that are within the range of each other, and the multi-hop communication to further optimize data sharing. In this scenario, the results also indicate that the system load, in terms of memory consumption, page faults, system calls, and context switches per second, increases. This is due to the greater amount of multi-hop communication among peers, which requires peers to maintain information about the available data around them.

\begin{table}[h]
\centering
\footnotesize
\begin{tabular}{|c|c|c|c|}
\hline
\textbf{Scenario} & \textbf{\begin{tabular}[c]{@{}c@{}}Download\\ Time (s)\end{tabular}} & \textbf{\begin{tabular}[c]{@{}c@{}}Number of\\Transmissions\end{tabular}} & \textbf{\begin{tabular}[c]{@{}c@{}}Memory \\ Overhead (MB)\end{tabular}} \\ \hline
1                 & 454                                                                            & 30,841                                                                            & 14.75                                                                                                                                                               \\ \hline
2                 & 418                                                                            & 24,243                                                                            & 14.65                                                                                                                                                                \\ \hline
3                 & 213                                                                            & 16,102                                                                            & 18.65                     							                                                                                  \\ \hline
\end{tabular}
\begin{tabular}[b]{|c|c|c|c|}
\hline
\textbf{Scenario} & \textbf{\begin{tabular}[c]{@{}c@{}}Context\\ Switches\end{tabular}} & \multicolumn{1}{c|}{\textbf{\begin{tabular}[c]{@{}c@{}}System\\ Calls\end{tabular}}} & \textbf{\begin{tabular}[c]{@{}c@{}}Page\\ Faults\end{tabular}} \\ \hline
1                 & 56,413                                                              & 214,313                                                                              & 4,742                                                          \\ \hline
2                 & 53,472                                                              & 202,542                                                                              & 4,683                                                          \\ \hline
3                 & 46,619                                                              & 186,548                                                                              & 4,274                                                          \\ \hline
\end{tabular}
\vspace{-0.1cm}
\caption{Real-world feasibility study results}
\label{Table:exp-results}
\end{table}

\vspace{-0.1cm}

\section{Conclusion \& Future Work}
\label{sec:conclusion}

In this paper, we presented DAPES, a data-centric design for off-the-grid peer-to-peer file sharing. DAPES offers mechanisms to maximize the utility of transmissions under intermittent connectivity and short-lived connections. It also achieves communication over multiple wireless hops by building short-lived knowledge about the data available around peers. 

While DAPES is off to a promising start, we plan to investigate several open issues in the future. First, we will compare DAPES with additional existing frameworks for off-the-grid file sharing and multi-hop communication. Second, we will conduct further real-world and simulation experiments, where peers share large numbers of file collections simultaneously under various mobility patterns. This will help us ``stress-test'' the scalability limits of our multi-hop communication design (e.g., amount of information that peers need to maintain) and our collision mitigation mechanism. Third, we will investigate the impact of having intermediate DAPES peers carry on behalf of others received Interests and data packets in their PIT and CS respectively. This direction will explore an off-the-grid file sharing approach that resembles more to Delay-Tolerant Networking (DTN)~\cite{fall2003delay} rather than MANET. Finally, previous work~\cite{li2001capacity, xu2002revealing} has demonstrated that the IEEE 802.11 MAC protocol may suffer from low throughput and high error rates in MANET scenarios. To this end, we plan to investigate the feasibility of data-centricity starting from the MAC layer of the network architecture all the way up to the application layer, and use DAPES as a driver example to investigate the impact of a data-centric MAC layer protocol~\cite{elbadry2018poster} on applications.





\bibliographystyle{plain}
\bibliography{reference} 

\end{document}